\documentclass[fleqn,usenatbib]{mnras}

\usepackage{newtxtext,newtxmath}

\usepackage[T1]{fontenc}

\DeclareRobustCommand{\VAN}[3]{#2}
\let\VANthebibliography\thebibliography
\def\thebibliography{\DeclareRobustCommand{\VAN}[3]{##3}\VANthebibliography}

\usepackage{graphicx}	
\usepackage{amsmath}	
\usepackage{verbatim}
\usepackage{ulem}

\newcommand{\hMsun}{\,$h^{-1}$\,M$_{\sun}$}
\newcommand{\hMpc}{\,$h^{-1}$\,Mpc}
\newcommand{\hkpc}{\,$h^{-1}$\,kpc}

\title[Dark matter halos in IDE models]{Dark Matter Halos in Interacting Dark Energy Models: Formation History, Density Profile, Spin and Shape}

\author[Y. Liu et al.]{
Yun Liu,$^{1}$
Shihong Liao,$^{2}$\thanks{E-mail: shihong.liao@helsinki.fi (SHL)}
Xiangkun Liu,$^{1}$\thanks{E-mail: liuxk@ynu.edu.cn (XKL)}
Jiajun Zhang,$^{3,4}$
Rui An$^{5}$
and Zuhui Fan$^{1}$\thanks{E-mail: zuhuifan@ynu.edu.cn (ZHF)}
\\
$^{1}$South-Western Institute for Astronomy Research, Yunnan University, Kunming 650500, China\\
$^{2}$Department of Physics, Gustaf Hällströmin katu 2, FI-00014, University of Helsinki, Finland\\
$^{3}$Center for Theoretical Physics of the Universe, Institute for Basic Science (IBS), Daejeon 34126, Korea\\
$^{4}$Shanghai Astronomical Observatory, Chinese Academy of Sciences, Shanghai 200030, China\\
$^{5}$Department of Physics and Astronomy, University of Southern California, Los Angeles, CA, 90089, USA
}

\date{Accepted 2022 January 21. Received 2021 December 22; in original form 2021 August 9}

\pubyear{2022}

\begin{document}
\label{firstpage}
\pagerange{\pageref{firstpage}--\pageref{lastpage}}
\maketitle

\begin{abstract}
The interacting dark energy (IDE) model, which considers the interaction between dark energy and dark matter, provides a natural mechanism to alleviate the coincidence problem and can also relieve the observational tensions under the $\Lambda$CDM model. Previous studies have put constraints on IDE models by observations of cosmic expansion history, cosmic microwave background and large-scale structures. However, these data are not yet enough to distinguish IDE models from $\Lambda$CDM effectively. Because the non-linear structure formation contains rich cosmological information, it can provide additional means to differentiate alternative models. In this paper, based on a set of $N$-body simulations for IDE models, we investigate the formation histories and properties of dark matter halos, and compare with their $\Lambda$CDM counterparts. For the model with dark matter decaying into dark energy and the parameters being the best-fit values from previous constraints, the structure formation is markedly slowed down, and the halos have systematically lower mass, looser internal structure, higher spin and anisotropy. This is inconsistent with the observed structure formation, and thus this model can be safely ruled out from the perspective of non-linear structure formation. Moreover, we find that the ratio of halo concentrations between IDE and $\Lambda$CDM counterparts depends sensitively on the interaction parameter and is independent of halo mass. This can act as a powerful probe to constrain IDE models. Our results concretely demonstrate that the interaction of the two dark components can affect the halo formation considerably, and therefore the constraints from non-linear structures are indispensable.
\end{abstract}




\section{Introduction}

The cosmological constant-cold dark matter ($\Lambda$CDM) model, which built on the framework of General Relativity and the hypotheses of dark energy and dark matter, is widely accepted so far as the standard cosmological model. It provides us physical scenarios of the evolution of Universe and the structure formation therein, which are supported by most of the cosmological observations today.

However, $\Lambda$CDM is still not the ultimate description for our Universe. From the theoretical aspect, the physical nature of dark energy and dark matter is not yet fully understood. There are long-standing fine-tuning and coincidence problems for the cosmological constant \citep{Weinberg1989RvMP...61....1W, Zlatev1999PhRvL..82..896Z}. Observationally, the $\Lambda$CDM model can explain cosmic large-scale structures nicely, but confronts acute problems at the dwarf-galaxy scale, namely the well-known missing satellites, core-cusp, and too-big-to-fail problems \citep[see][for a review]{Bullock2017ARA&A..55..343B}. In recent years, with the rapid observational developments, a few inconsistencies between different observations under $\Lambda$CDM have emerged and become increasingly significant. The most notable one is the $H_0$ tension with the Hubble constant $H_0$ derived from Planck cosmic microwave background (CMB) observations $\sim 5\sigma$ away from the ones measured from standard candles and the lensing time-delay observations \citep{Riess2020NatRP...2...10R,Riess2019ApJ...876...85R,Wong2020MNRAS.498.1420W}. The other less significant one is the $S_8$ tension where there is a $1-2\sigma$ discordance between the value of $S_8=\sigma_8(\Omega_{\rm m})^{0.5}$ constrained from Planck CMB and that from weak lensing measurements \citep{Hikage2019PASJ...71...43H,Hildebrandt2020A&A...633A..69H,Asgari2021A&A...645A.104A,Amon2021arXiv210513543A}.

Given the problems, different non-standard cosmological scenarios have been proposed. Among them, the interacting dark energy (IDE) models have been regarded as physically 
well motivated ones. In this framework, the two dark components that dominate the current evolution of the Universe are not independent but interact with each other. Thus their densities coevolve, giving rise to a natural explanation for the coincidence problem \citep{Amendola2000,Amendola2003,Amendola2006,Pavon2005,Del2008,Boeher2008,Olivares2006,Chen2008}. Furthermore, it has been testified that the IDE model with appropriate parameters is also effective to relieve the discordances under a $\Lambda$CDM framework that are mentioned before \citep{Ferreira2017PhRvD..95d3520F,An2018JCAP...02..038A}. There have been extensive studies for IDE models, both from theoretical aspects \citep{He2009PhLB..671..139H, He2009JCAP...07..030H, He2010JCAP...12..022H,D'Amico2016PhRvD..94j3526D} and from observational constraints \citep{Costa2017JCAP...01..028C,Zhang2019ApJ...875L..11Z,Cheng2020PhRvD.102d3517C,Kang2021PDU....3100784K}. See \citet{Wang2016RPPh...79i6901W} and \citet{Bolotin2015IJMPD..2430007B} for detailed reviews of IDE models.

It is noted that the current constraints on IDE models are mainly from cosmic expansion history, CMB and the structure formation on large scales. Their effects on non-linear structure formation have not been fully investigated. Physically, the interaction between dark energy and dark matter induces their densities change in addition to that due to cosmic expansion. Furthermore, the accelerating force on dark matter particles is also altered. We therefore expect that the non-linear structure formation, in particular the properties of dark matter halos, can be considerably different in IDE models in comparison with their $\Lambda$CDM counterparts, and thus can provide valuable information in constraining IDE models. The non-linear matter power spectra of IDE models have been studied in the recent years \citep{Baldi2011MNRAS.414..116B,baldi2012mnras,casas2016jcap}. In \citet{Carlesi2014MNRAS.439.2943C,Carlesi2014MNRAS.439.2958C}, they analysed the properties of dark matter halos in different cosmic web environments at redshift $z=0$ from hydrodynamical simulations of coupled and uncoupled quintessence models. The gas properties in clusters of galaxies were also investigated. Their findings show that the differences of halo properties, such as the spin and the concentration, between their considered models depend sensitively on the surrounding environment of halos. The initial attempt to study the halo evolution in the coupled dark sector model was done in \citet{jibrail2020mnras}. However, the details of halo formation history and its properties are still lacking study.

Recently, a fully self-consistent $N$-body simulation pipeline for IDE models, \texttt{ME-GADGET}, was developed by \cite{Zhang2018PhRvD..98j3530Z}, which enables us to trace the non-linear structure formation in IDE models accurately. Therefore, in this paper, based on a set of simulations from \texttt{ME-GADGET}, we analyse systematically the formation history and properties of dark matter halos in IDE models, 
including mass function, density profile, spin and shape.

The paper is organized as follows. We first introduce the phenomenological IDE models (section~\ref{sec:Models}) and our $N$-body simulations (section~\ref{sec:Simulations}), then, we describe the dark matter halo catalogues used in this study (section~\ref{sec:Samples}). We present our results in section~\ref{sec:Results}. Summary and discussions are presented in section~\ref{sec:Conclusions}.

\section{Methods}

\subsection{Phenomenological IDE models}
\label{sec:Models}

\begin{table}
	\centering
	\begin{tabular}{c c c}
		\hline
		Model & $Q$ & $w_{\mathrm{d}}$\\
		\hline
		IDE1 & $3\mathcal{H}\xi_{2}\rho_{\mathrm{d}}$ & $-1<w_{\mathrm{d}}<-1/3$\\
		IDE2 & $3\mathcal{H}\xi_{2}\rho_{\mathrm{d}}$ & $w_{\mathrm{d}}<-1$\\
		IDE3 & $3\mathcal{H}\xi_{1}\rho_{\mathrm{c}}$ & $w_{\mathrm{d}}<-1$\\
		IDE4 & $3\mathcal{H}\xi(\rho_{\mathrm{d}}+\rho_{\mathrm{c}})$ & $w_{\mathrm{d}}<-1$\\
		\hline
	\end{tabular}
	\caption{IDE models that have been studied.}
	\label{tab:IDE models}
\end{table}

In the $\Lambda$CDM model, dark energy and dark matter evolve independently with their energy densities conserved separately. In contrast, in the IDE models, due to the additional postulated interaction, only the total energy density of the dark sector is conserved, and the energy-momentum tensor of the dark sector, $T^{\mu\nu}_{(i)}$, satisfies
\begin{equation}
    \sum_{i}\nabla_{\mu}T^{\mu\nu}_{(i)}=0,
	\label{eq:IDEtensor}
\end{equation}
where $i$ represents dark energy (d) or dark matter (c), and Einstein's summation convention has been adopted. Using the energy-momentum tensor of a perfect fluid, and adopting a flat Friedmann-Lemaître-Robertson-Walker metric, we can deduce the conservation equations in the background level \citep{He2009PhLB..671..139H}
\begin{equation}
    \begin{split}
        \rho^{\prime}_{\mathrm{d}}+3\mathcal{H}(1+w_{\mathrm{d}})\rho_{\mathrm{d}}&=-Q\\
        \rho^{\prime}_{\mathrm{c}}+3\mathcal{H}\rho_{\mathrm{c}}&=Q,
        \label{eq:IDEconservation}
    \end{split}
\end{equation}
where $\rho_{\rm d}$ and $\rho_{\rm c}$ are the energy densities of dark energy and dark matter respectively, $w_{\mathrm{d}}\equiv p_{\mathrm{d}}/\rho_{\mathrm{d}}$ is the equation of state for dark energy with $p_{\mathrm{d}}$ being the pressure, $\mathcal{H}=a^{\prime}/a$ is the Hubble parameter with $a$ being the cosmic scale factor, the prime symbol represents the derivative with respect to the conformal time, and $Q$ is the interaction kernel. Here, $Q>0$ reflects that the energy flows from dark energy to dark matter while $Q<0$ the opposite.

In the current phenomenological IDE models, $Q$ is commonly assumed to be a function of energy density $\rho$ and of time $\mathcal{H}^{-1}$, i.e.
\begin{equation}
    Q=Q(\rho,\mathcal{H}^{-1}),
	\label{eq:QQ}
\end{equation}
and the Taylor expansion at the first order can be written as \citep{Wang2016RPPh...79i6901W}
\begin{equation}
    Q=3\mathcal{H}(\xi_{1}\rho_{\mathrm{c}}+\xi_{2}\rho_{\mathrm{d}}),
	\label{eq:Q}
\end{equation}
where $\xi_{1}$ and $\xi_{2}$ are free parameters. Table~\ref{tab:IDE models} shows the models that have been studied in practice. The requirement on $w_{\rm d}$ parameter is set based on the stability behavior of perturbations \citep{He2009PhLB..671..139H}.

In \citet{Costa2017JCAP...01..028C}, constraints on the models in Table~\ref{tab:IDE models} are derived using Planck CMB, BOSS BAO, type $\mathrm{\uppercase\expandafter{\romannumeral1}}$a supernova and Hubble constant observations. While tight constraints are obtained for IDE3 and IDE4, only weak ones are achievable for IDE1 and IDE2. This is because the interactions in IDE1 and IDE2 are proportional to dark energy density, which came to be dominant only recently. Such an interaction has weak impacts on CMB anisotropy and cosmic expansion history, and thus cannot be tightly constrained from the corresponding observations. However, the non-linear structure formation in these two models can be significantly different from that of $\Lambda$CDM model. Therefore tighter constraints on IDE1 and IDE2 are expected by employing probes related to non-linear structures, such as the abundance of clusters of galaxies and weak lensing peak statistics. This is the motivation of our study here. Based on a set of IDE $N$-body simulations, we will perform systematic analyses on the non-linear structure formation in different models.

\subsection{\textit{N}-body simulations for IDE models}
\label{sec:Simulations}

\begin{table}
	\centering
	\begin{tabular}{c c c c c c}
		\hline
		Parameter & $\Lambda$CDM & IDE1 & IDE2 & IDE1$^\prime$ & IDE2$^\prime$\\
		\hline
		$\ln{(10^{10}A_{\mathrm{s}})}$ & 3.094 & 3.099 & 3.097 & 3.0965 & 3.0955\\
		$n_{\mathrm{s}}$ & 0.9645 & 0.9645 & 0.9643 & 0.9645 & 0.9644\\
		$w_{\mathrm{d}}$ & -1 & -0.9191 & -1.088 & -0.95955 & -1.044\\
		$\xi_{2}$ & 0 & -0.1107 & 0.05219 & -0.05535 & 0.026095\\
		$h$ & 0.6727 & 0.6818 & 0.6835 & 0.67725 & 0.6808\\
		$\Omega_{\mathrm{d}}$ & 0.6844 & 0.7817 & 0.6631 & 0.7345 & 0.6769\\
		$\Omega_{\mathrm{m}}$ & 0.3156 & 0.2182 & 0.3368 & 0.2654 & 0.3230\\
		$\Omega_{\mathrm{r}}$ & 0 & 0.0001 & 0.0001 & 0.0001 & 0.0001\\
		\hline
	\end{tabular}
	\caption{Cosmological parameters for $\Lambda$CDM and IDE simulations. The $\Lambda$CDM parameters are from the Planck 2015 results \citep{Planck2016A&A...594A..13P}, while the parameters for IDE1 and IDE2 are the best-fit values from the constraints using Planck CMB + BAO + SN$\mathrm{\uppercase\expandafter{\romannumeral1}}$a + $H_0$ by \citet{Costa2017JCAP...01..028C}. For the IDE1$^\prime$ and IDE2$^\prime$, their parameters are chosen to be at the midpoints between $\Lambda$CDM and the corresponding best-fit IDE models.}
	\label{tab:Parameters}
\end{table}

\begin{figure}
	\includegraphics[width=\columnwidth]{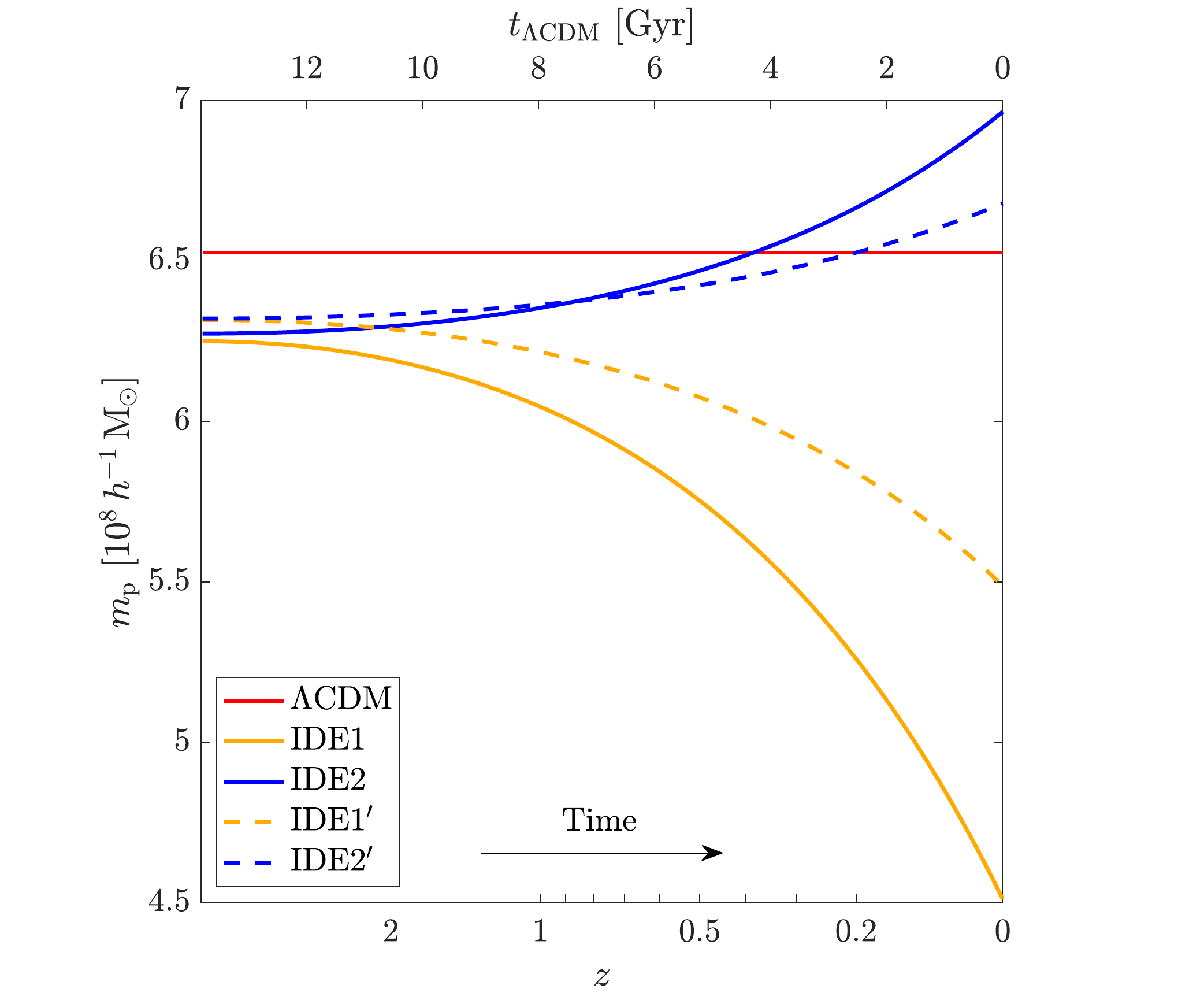}
    \caption{Evolution of particle masses in $\Lambda$CDM (red), IDE1 (yellow solid), IDE2 (blue solid), IDE1$^\prime$ (yellow dashed) and IDE2$^\prime$ (blue dashed) simulations. The lower horizontal axis shows the redshift, while the upper horizontal axis gives the corresponding looking-back time in $\Lambda$CDM.}
    \label{fig:Particle mass}
\end{figure}

\begin{figure*}
	\includegraphics[width=2\columnwidth]{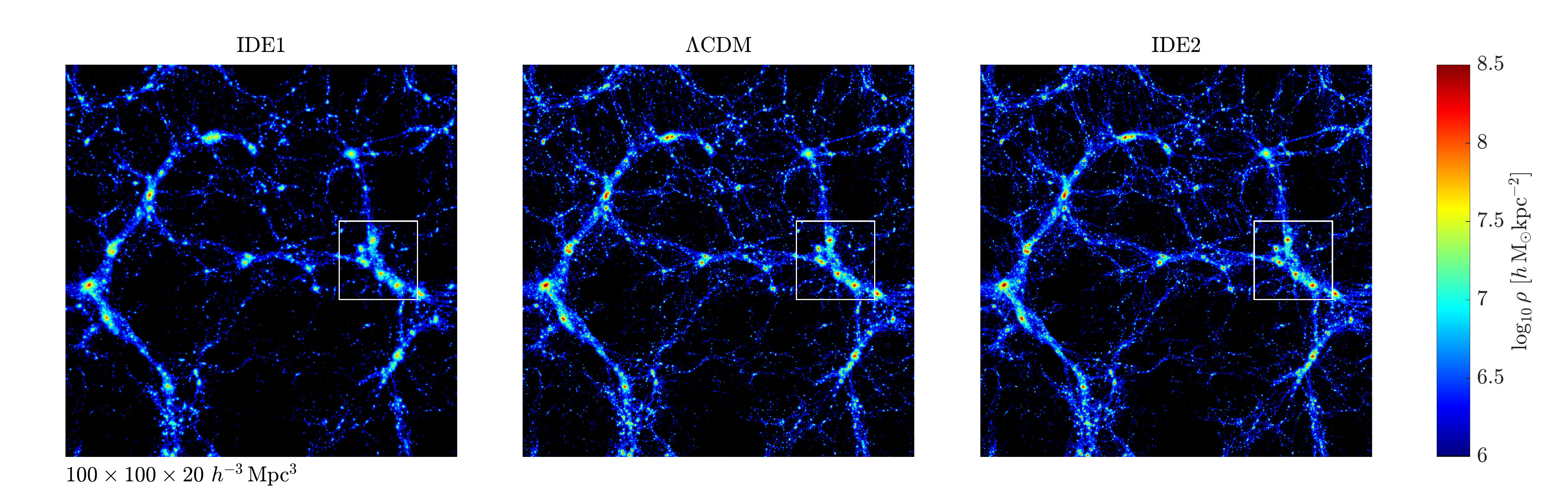}
	\includegraphics[width=2\columnwidth]{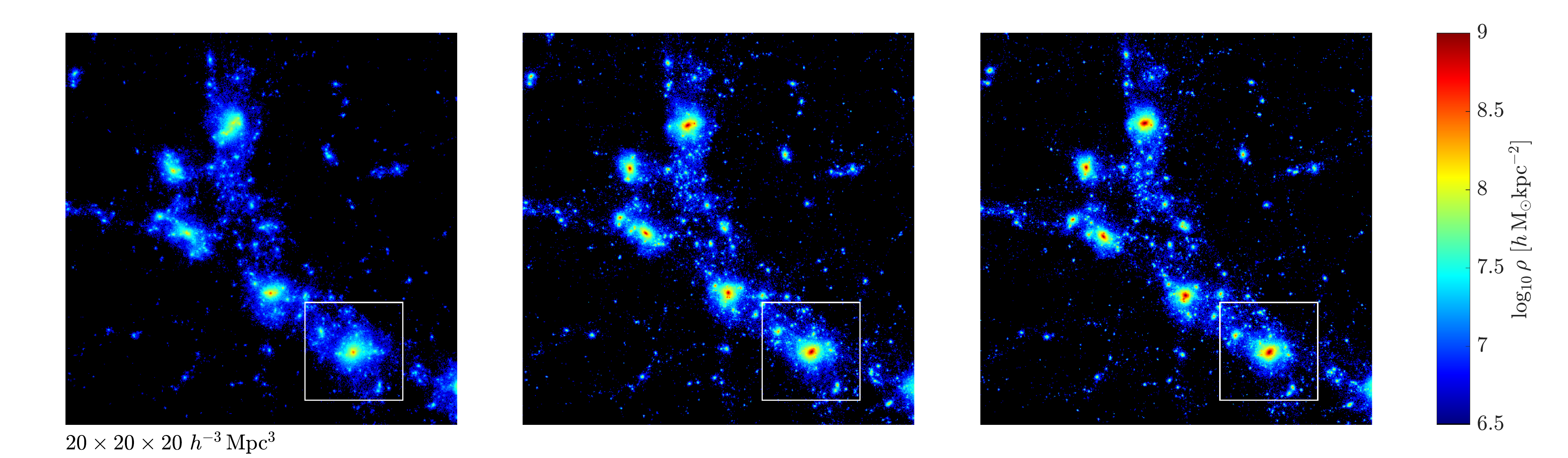}
	\includegraphics[width=2\columnwidth]{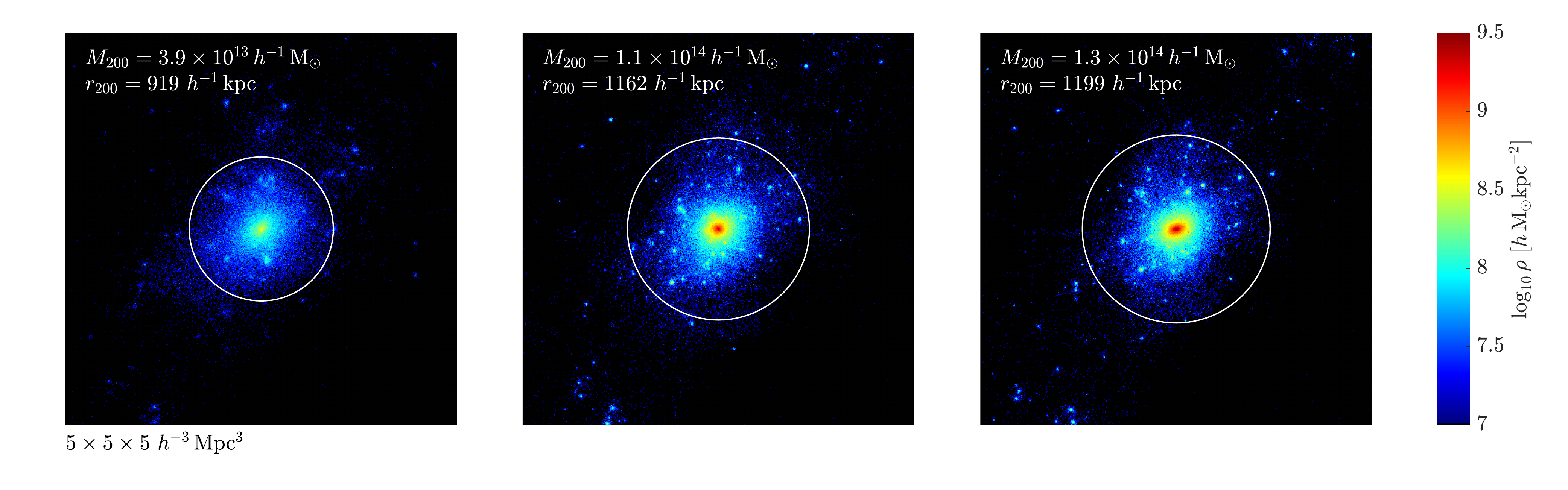}
    \caption{Contrasts on the $z=0$ cosmic structures in IDE1 (left column), $\Lambda$CDM (middle column) and IDE2 (right column) models at different scales. Top row: 2-D projections of the matter density distribution of a simulation slice with a side length of {$100$\hMpc} and a thickness of {$20$\hMpc}. Middle row: zoom-in plots of the high-density regions which are marked with the white squares in the top panels. Bottom row: zoom-in plots of a $\Lambda$CDM halo and its IDE counterparts which are marked with the white squares in the middle panels. The white circles in the bottom panels mark the halo virial radii $r_{200}$. In each panel, the projected density is shown in logarithmic scale, and the color bar for each row is shown on the right.}
    \label{fig:LSS}
\end{figure*}

To perform the simulations for IDE models, we have employed the \texttt{ME-GADGET} code \citep{Zhang2018PhRvD..98j3530Z} which is based on the public TreePM \texttt{GADGET-2} code \citep{Springel2005MNRAS.364.1105S} with the following critical modifications. First, because of the interactions between dark energy and dark matter, the IDE1 and IDE2 models have evolving matter densities different from that of the $\Lambda$CDM model. It is given by
 \begin{equation}
    \rho_{\mathrm{m}}(a)=\frac{3H_{0}^{2}}{8\pi G}\left[\left(\Omega_{\mathrm{m}}+\frac{\xi_{2}\Omega_{\mathrm{d}}}{w_{\mathrm{d}}+\xi_{2}}\right)a^{-3}-\frac{\xi_{2}\Omega_{\mathrm{d}}}{w_{\mathrm{d}}+\xi_{2}}a^{-3(1+w_{\mathrm{d}}+\xi_{2})}\right],
	\label{eq:rho_m(a)}
\end{equation}
where $H_0=100h\,{\rm km}\,{\rm s}^{-1}\,{\rm Mpc}^{-1}$ is the Hubble constant, and $G$ is the gravitational constant. As a result, the mass of the simulated particles, $m_{\mathrm{p}}(a)$, is changing with the cosmic scale factor $a$. Second, the evolution of Hubble parameter, $H(a)$, is also modified accordingly. Third, due to the interaction between the two dark components, the perturbation of dark energy induces an additional force on dark matter particles, leading to the modified Poisson equation. Finally, there is also an extra acceleration for dark matter particles that is proportional to the interaction parameter and the particle velocity. We refer the readers to \citet{Zhang2018PhRvD..98j3530Z} for further code details.

To generate self-consistent initial conditions, we first use the capacity constrained Voronoi tessellation method \citep{Liao2018MNRAS.481.3750L,Zhang2021MNRAS.507.6161Z} to produce a uniform and isotropic particle distribution. We then modify the \texttt{\textsc{camb}} code \citep{Lewis2002PhRvD..66j3511L} to calculate the linear matter power spectra for IDE models and use the \texttt{\textsc{2LPTic}} code \citep{Crocce2006MNRAS.373..369C} to generate the initial perturbed position and velocity for the particles. See \citet{Zhang2018PhRvD..98j3530Z} for details.

In our studies here, we consider five sets of simulations, $\Lambda$CDM, IDE1, IDE2, IDE1$^\prime$ and IDE2$^\prime$. The cosmological parameters adopted are shown in Table~\ref{tab:Parameters}. The ones for the $\Lambda$CDM simulation are consistent with the Planck 2015 results \citep{Planck2016A&A...594A..13P}. The IDE1 and IDE2 employ the best-fit parameters constrained by \citet{Costa2017JCAP...01..028C}. We note that for utilizing the redshift space distortion (RSD) data into constraints on IDE models, consistent treatments using IDE templates to extract RSD information from galaxy redshift-space distributions are needed. These are still lacking in the current RSD analyses. We therefore choose to use the fitting parameters in Table 4 and Table 5 of \citet{Costa2017JCAP...01..028C} for IDE1 and IDE2, respectively, derived from Planck CMB + BAO + SN$\mathrm{\uppercase\expandafter{\romannumeral1}}$a + $H_0$ without RSD. Additionally, we run two more IDE simulations, IDE1$^\prime$ and IDE2$^\prime$, with their parameters chosen to be the midpoint values between $\Lambda$CDM and best-fit IDE1 and IDE2, respectively, which are well within the allowed parameter ranges from \citet{Costa2017JCAP...01..028C}.

Each simulation evolves $512^3$ particles from $z=127$ to $0$ in a periodic box with a side length of {100\hMpc}. Each simulation has 139 outputs in total, and the time interval for snapshots is $\sim 0.1$ Gyr. The comoving gravitational softening length used in our simulations is {4\hkpc}, which is $\sim1/50$ of the mean inter-particle separation.

The particle mass in the $\Lambda$CDM simulation is $6.53\times10^{8}$\hMsun. For IDE models, the particle mass is changing with time due to the interaction between dark energy and dark matter. Figure~\ref{fig:Particle mass} shows the particle mass vs. redshift in different simulations, where the upper horizontal axis is the corresponding cosmic time in $\Lambda$CDM. For IDE1, dark matter decays into dark energy and, thus, the particle mass decreases from {$6.25\times10^{8}$\hMsun} at $z=127$ to {$4.51\times10^{8}$\hMsun} at $z=0$. For IDE2, the energy flows from dark energy into dark matter, and the particle mass increases from {$6.27\times10^{8}$\hMsun} to $6.96\times10^{8}$\hMsun. Similar trends are seen for IDE1$^\prime$ and IDE2$^\prime$, but the changes are milder because of the smaller values of $|\xi_2|$.

We have used the same random seed to generate the initial conditions for the five simulations and, therefore, are able to conduct direct comparisons of the large-scale structures. Figure~\ref{fig:LSS} presents the cosmic matter density distributions of different scales at $z=0$ from our simulations. From the top panels, we can see that the large-scale structures ($\gtrsim 10$\hMpc) in IDE and $\Lambda$CDM models are fairly similar. However, when we zoom into smaller scales as shown in the middle and bottom panels, we can clearly see that the IDE1 (IDE2) structures tend to have lower (higher) central peak densities comparing to the $\Lambda$CDM counterparts. This illustrates that the differences between IDE and $\Lambda$CDM models are more enhanced at small-scale virialized structures, and it hints us that studying and quantifying the halo properties at low redshifts in IDE models can help to put tighter constraints on these models.

\subsection{Dark matter halo catalogues}
\label{sec:Samples}

\begin{figure}
	\includegraphics[width=\columnwidth]{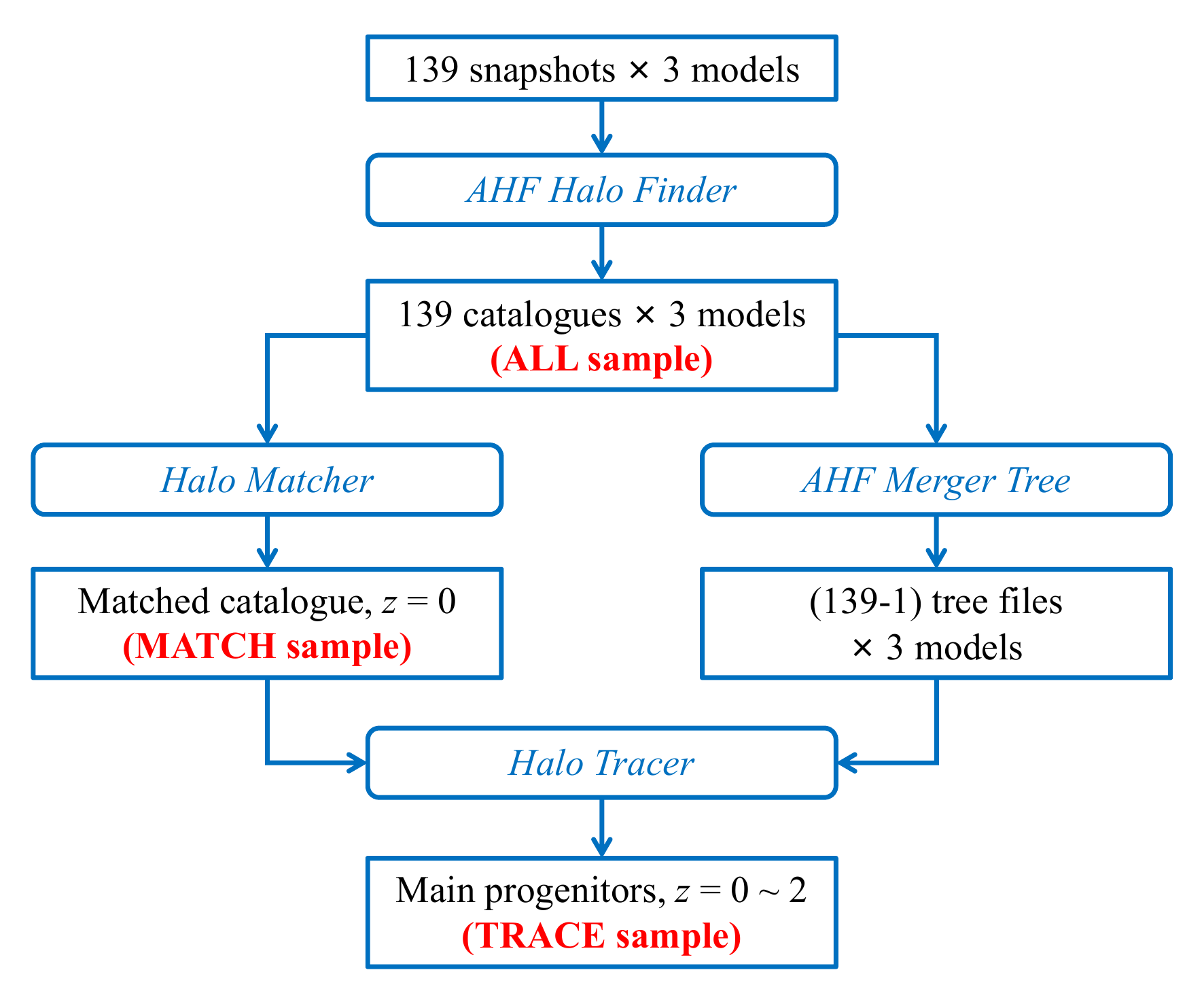}
    \caption{The pipeline to generate the ALL, MATCH and TRACE halo samples based on the $\Lambda$CDM, IDE1 and IDE2 simulations.}
    \label{fig:pipeline}
\end{figure}

Based on the snapshot outputs of the $\Lambda$CDM, IDE1, IDE2, IDE1$^\prime$ and IDE2$^\prime$ simulations, we construct different dark matter halo samples for analyses, namely ALL, MATCH and TRACE samples. Figure~\ref{fig:pipeline} summarizes our construction pipeline considering the $\Lambda$CDM, IDE1 and IDE2 models as an illustration. We describe in detail the steps in this section. The same procedures are also performed parallelly for the case of IDE1$^\prime$ and IDE2$^\prime$.

\subsubsection{ALL sample}
\label{sec:ALL}

We employ the AHF halo finder \citep{Gill2004MNRAS.351..399G,Knollmann2009ApJS..182..608K}, which works based on the spherical overdensity (SO) algorithm, to identify dark matter halos from simulation snapshot outputs. We adopt the centre of mass position as the halo centre, and define the halo virial radius, $r_{200}$, as the radius within which the mean density is 200 times the cosmic matter density, $\rho_{\mathrm{m}}(z)$, at the corresponding redshift. For IDE models, we modify the related calculations in the \texttt{AHF} code to take into account the time evolution of the cosmic mean matter density (Equation~\ref{eq:rho_m(a)}). We run the modified \texttt{AHF} code for all snapshots of the $\Lambda$CDM, IDE1 and IDE2 simulations from which $139\times3$ dark matter halo catalogues are obtained, we refer to them as ALL sample hereafter. The minimum particle number used in our halo identification is 50. However, to estimate the halo properties robustly, unless otherwise specified, we only use halos with {$M_{200}>10^{11}$\hMsun} (i.e. those have at least $\sim154$ particles in the $\Lambda$CDM simulation) in our analyses.

\subsubsection{MATCH sample}
\label{sec:MATCH}

Because our simulations for different models start with the same set of initial random seeds, we can match the halo counterparts in different models and perform the one-to-one comparison. For that, we construct the matched halo sample at $z=0$ by using the \texttt{AHF Merger Tree} code to cross-match the IDs of halo particles in different runs. Two halos from different models are considered as a prospective counterpart pair if they maximize each other's merit function
\begin{equation}
    \mathcal{M}_{\mathrm{AB}}=\frac{N_{\mathrm{A \cap B}}^{2}}{N_{\mathrm{A}} \cdot N_{\mathrm{B}}},
	\label{eq:Merit function}
\end{equation}
where $N_{\mathrm{A}}$ and $N_{\mathrm{B}}$ are the particle numbers of two halos in model A and B respectively, and $N_{\mathrm{A \cap B}}$ is the number of their common particles (i.e. particles with the same IDs).

Specifically, for all $z=0$ halos which are more massive than {$10^{11}$\hMsun} in the ALL sample, we match the $\Lambda$CDM halos with IDE1 and IDE2 halos respectively, and only keep those halos which have both IDE1 and IDE2 counterparts. These triplets serve as our initial matched halo sample. Then, the following selections are carried out:
\begin{enumerate}
 \item Halos in the IDE1 model have systematically lower masses than their $\Lambda$CDM counterparts, which will be shown in Section~\ref{sec:Results}. To ensure the completeness of the matched sample, we therefore only consider the $\Lambda$CDM halos with {$M_{200}\geq10^{12}$\hMsun} so that their counterparts in IDE models are all more massive than {$10^{11}$\hMsun}.
 \item Because of the mass resolution limitation of our simulations, we mainly focus on the properties of host halos in this study without analysing subhalos. Thus we remove a halo triplet from the initial matched sample if any of the three counterparts is 
a subhalo.
 \item We pick out the halo triplets which have both $\mathcal{M}_{\Lambda \mathrm{CDM-IDE1}}$ and $\mathcal{M}_{\Lambda \mathrm{CDM-IDE2}}$ greater than 0.25. From the definition of the merit function, $\mathcal{M}_{\mathrm{AB}}>0.25$ means that at least half of the particles in halo A (or halo B) are common particles.
\end{enumerate}

After applying the selections above, out of 4174 $\Lambda$CDM host halos more massive than {$10^{12}$\hMsun} at $z=0$, there are 3763 ones have reliable counterparts in both IDE1 and IDE2 simulations. They form the MATCH sample. The bottom row of Figure~\ref{fig:LSS} shows an example of a $\Lambda$CDM halo and its IDE counterparts.

\subsubsection{TRACE sample}
\label{sec:TRACE}

To study the halo formation history, we use the \texttt{AHF Merger Tree} code to construct the merger trees for all $z=0$ halos in the MATCH sample. Specifically, for a target halo in a certain snapshot, the \texttt{AHF Merger Tree} code searches for its progenitors in the previous adjacent snapshot by cross-matching halo particle IDs. If a candidate halo in the previous adjacent snapshot shares at least 10 common particles with the target halo, it is included as one of the progenitors of the target halo. The main progenitor is defined as the one which has the maximum merit function with the target halo.

To construct stable trees, the \texttt{AHF Merger Tree} code has the function to skip snapshot(s) when a halo is going through complicated environment and to cope with temporary large fluctuation in mass \citep[see][for more discussions]{Srisawat2013MNRAS.436..150S}. In this study, we require that if a target halo cannot find any progenitor in the adjacent snapshot or the mass ratio between the target halo and the progenitor candidate is larger than 2, the code will ignore this snapshot and continue to search for the credible progenitors in high-redshift snapshots.

In total, there are 3010 halo triplets at $z=0$ from the MATCH sample whose merging histories can be traced back to $z>2$ in all three models. We name them as the TRACE sample.

\section{Results}
\label{sec:Results}

With the halo catalogues described above, we study the abundance, formation history and internal properties of dark matter halos in the IDE models, and compare them with the $\Lambda$CDM counterparts. We note that the five models considered here do not have exactly the same $H_0$ value, but the differences are small (i.e. $<2\%$, see Table~\ref{tab:Parameters}). We therefore use the conventional units of {\hMsun} and {\hMpc} (\hkpc) for mass and length in model comparisons, and ignore the differences in $H_0$.

\subsection{Halo mass function}
\label{sec:Mass function}

\begin{figure*}
	\includegraphics[width=\columnwidth]{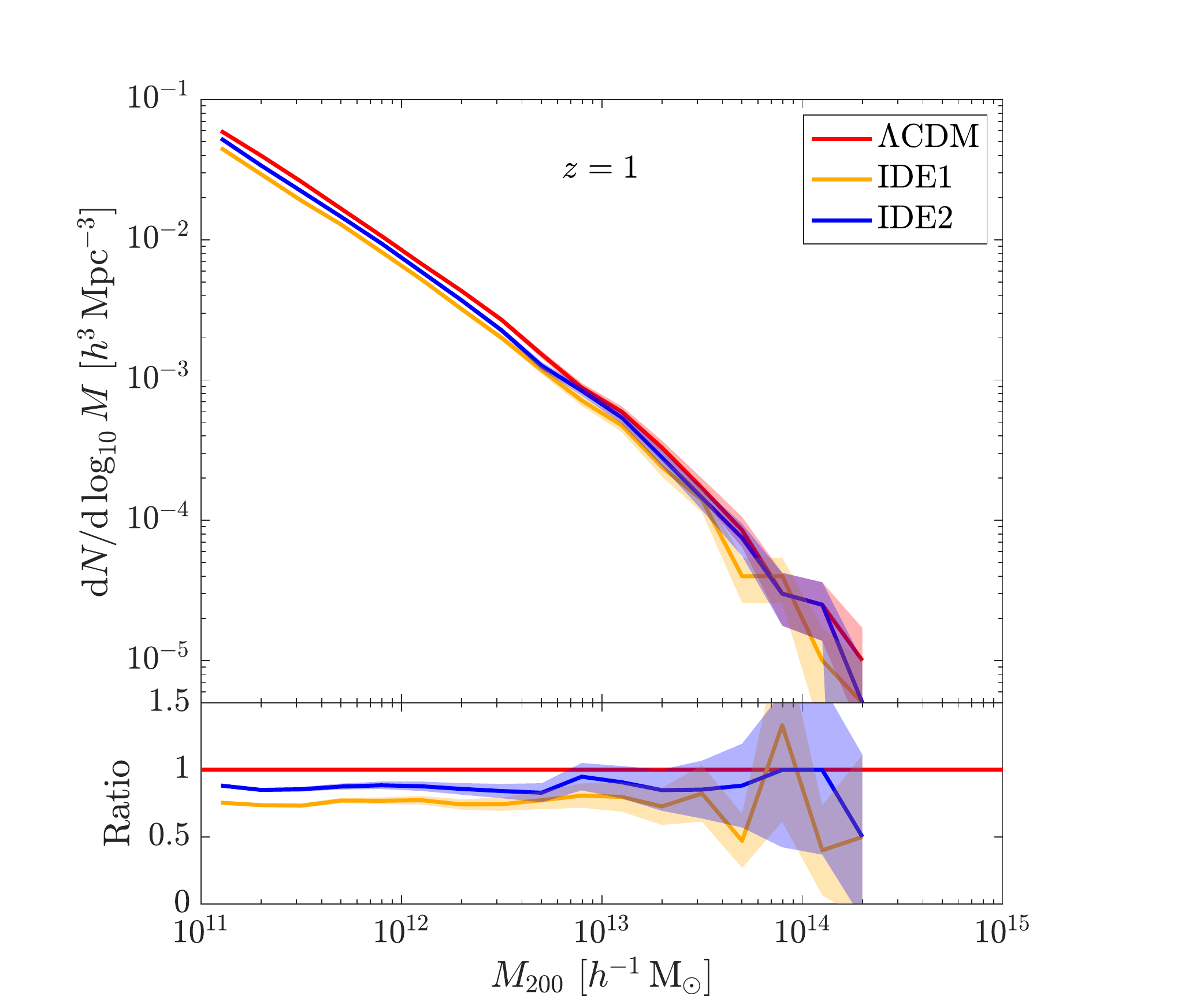}
	\includegraphics[width=\columnwidth]{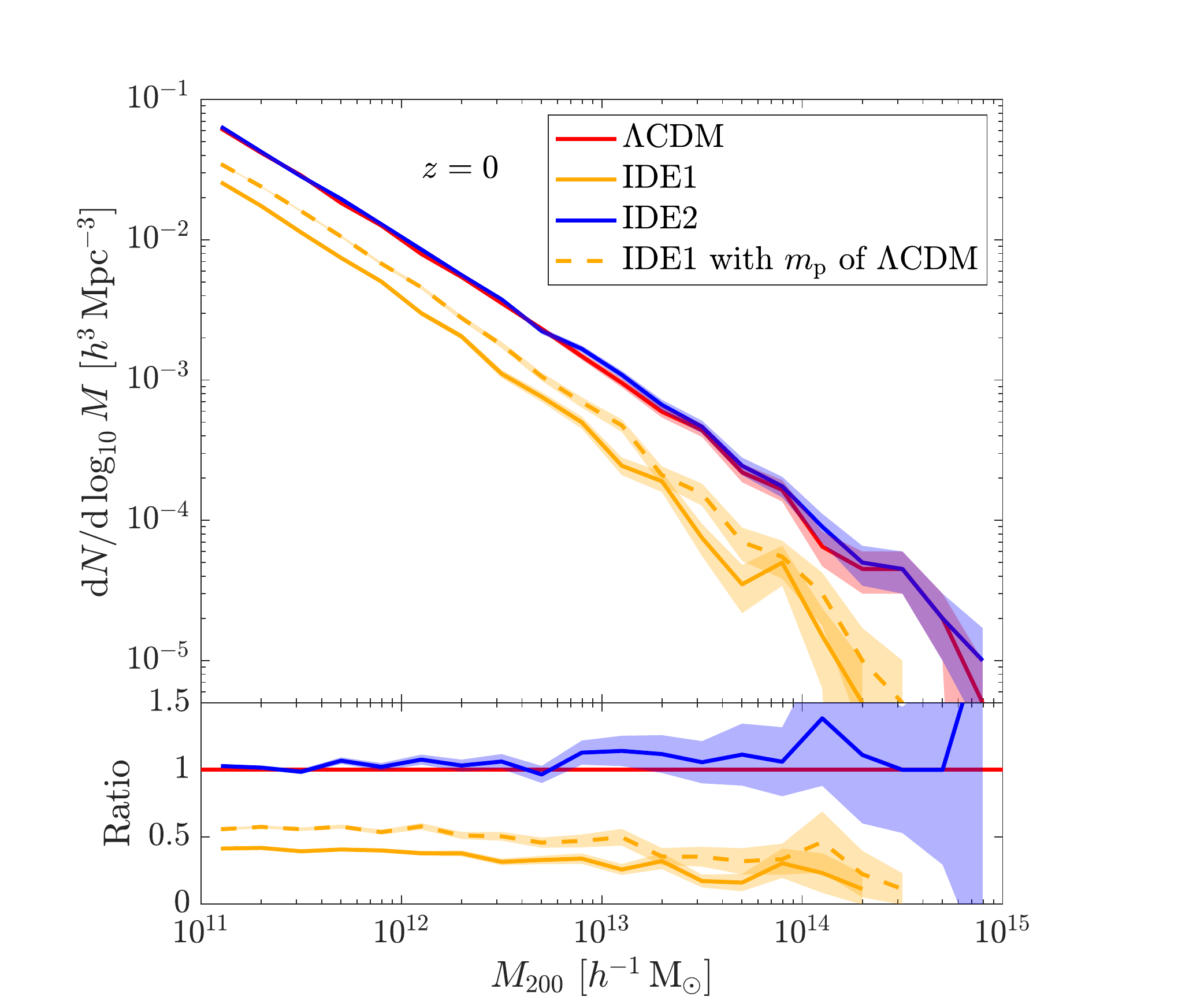}
    \caption{Halo mass functions at $z=1$ (left) and $z=0$ (right) in $\Lambda$CDM (red), IDE1 (yellow solid) and IDE2 (blue) models, respectively. The upper panels show the differential halo mass functions, while the lower panels show the ratios with respect to $\Lambda$CDM. The shaded region associated with each line represents the Poisson error. In the right panel, the yellow dashed lines plot the rescaled IDE1 halo mass function (see the text for detailed discussions).}
    \label{fig:MF}
\end{figure*}

We first analyse the halo mass function, i.e. the comoving number density of dark matter halos as a function of halo mass and redshift, $n(M,z)$, for different models. As one of the most important quantities describing the non-linear structure formation, the halo mass function carries rich cosmological information and is the foundation to extract cosmological parameter constraints from observations of galaxy cluster number counts \citep{Abbott2020PhRvD.102b3509A,Allen2011ARA&A..49..409A} and weak lensing peak statistics \citep{Hamana2004MNRAS.350..893H,Fan2010ApJ...719.1408F}. It also plays an essential role in building halo models \citep{Cooray2002PhR...372....1C,Skibba2009MNRAS.392.1080S} and the semi-analytic models of galaxy formation \citep{Conroy2006ApJ...647..201C}.

For $\Lambda$CDM model, there have been extensive studies on the halo mass function \citep[e.g.][]{Press1974ApJ...187..425P,Bond1991ApJ...379..440B,Sheth2001MNRAS.323....1S,Jenkins2001MNRAS.321..372J,Reed2003MNRAS.346..565R,Tinker2008ApJ...688..709T,Watson2013MNRAS.433.1230W}. In the IDE framework, \citet{He2009JCAP...07..030H} calculated the linear growth of perturbations, and subsequently \citet{He2010JCAP...12..022H} studied the spherical collapse systems.

Here we analyse the halo mass functions of different models based on the ALL sample, as shown in Figure~\ref{fig:MF} for $\Lambda$CDM, IDE1 and IDE2. In order to investigate the evolution, besides the halo mass functions at $z=0$ (right panel), we also measure the results at $z=1$ (left panel) when the dark energy-dark matter interactions are still weak. At $z=1$, both IDE models produce systematically less halos than $\Lambda$CDM with the ratio of $\sim0.76$ and $\sim0.87$ for IDE1 and IDE2, respectively. This is mainly because of the lower $\rho_{\rm m}(z)$ at high redshift for the two models than that of $\Lambda$CDM (see Figure~\ref{fig:Particle mass} and Equation~\ref{eq:rho_m(a)}).

At $z=0$, we see that the halo mass function of IDE1 model is markedly lower than that of $\Lambda$CDM over the whole considered mass range, and the difference increases at the high mass end. For $M_{200}\approx10^{11}$\hMsun, the ratio between IDE1 and $\Lambda$CDM is $\sim0.41$, while for {$M_{200}=10^{14}$\hMsun} the ratio reduces to $\sim0.23$. For the considered IDE models, the interaction between dark energy and dark matter is proportional to the dark energy density. When dark energy becomes a dominant component at lower redshift, the interaction gets stronger. For IDE1, dark matter then decays into dark energy more effectively, which suppresses the growth of dark matter structures and even reduces the halo mass (see Section \ref{sec:Formation history} for detailed discussions). We therefore observe a significantly lower halo mass function at $z=0$ in the IDE1 model.

In contrast to the IDE1 model, the IDE2 halo mass function at $z=0$ is slightly higher than the $\Lambda$CDM one, with a ratio of $\sim1.1$ on average. For IDE2, when dark energy dominates at low redshift, more energy flows into dark matter. This increases the dark matter content and thus accelerates the growth of dark matter halos. Therefore, although the IDE2 halo mass function is lower at $z=1$, it gradually catches up at lower redshift and surpasses that of the $\Lambda$CDM model at $z=0$.

As seen from Figure~\ref{fig:Particle mass}, because of the interaction, the dark matter particle mass in our IDE simulations is changing with time, and decreases or increases for IDE1 and IDE2, respectively. To see if the difference in the halo mass functions can be fully attributed to the particle mass difference, we rescale the mass of each IDE1 halo at $z=0$ by a factor of $m_{\rm p, \Lambda CDM}/m_{\rm p, IDE1}$. This is shown as the yellow dashed line in the right panel of Figure~\ref{fig:MF}. We can clearly see that the rescaled IDE1 halo mass function is still much lower than the $\Lambda$CDM one. Quantitatively, on average between $M_{200}=10^{11}\sim10^{13}$\hMsun, the relative difference between the IDE1 and $\Lambda$CDM halo mass functions is $1-{\rm Ratio}\approx0.62$, and the relative difference between the rescaled IDE1 and $\Lambda$CDM ones is $1-{\rm Ratio}\approx0.47$. Thus the difference in particle mass only contributes $\sim 15\%$ to the difference between IDE1 and $\Lambda$CDM halo mass functions. The dominant effect is from the suppression of the structure growth in IDE1 model because dark matter keeps decaying into dark energy.

For IDE1$^\prime$ and IDE2$^\prime$, the halo mass functions have similar behaviors as that of IDE1 and IDE2, respectively. But the differences to the $\Lambda$CDM model are smaller because of the smaller interaction parameter $|\xi_2|$.

\subsection{Halo formation history}
\label{sec:Formation history}

\begin{figure}
    \includegraphics[width=\columnwidth]{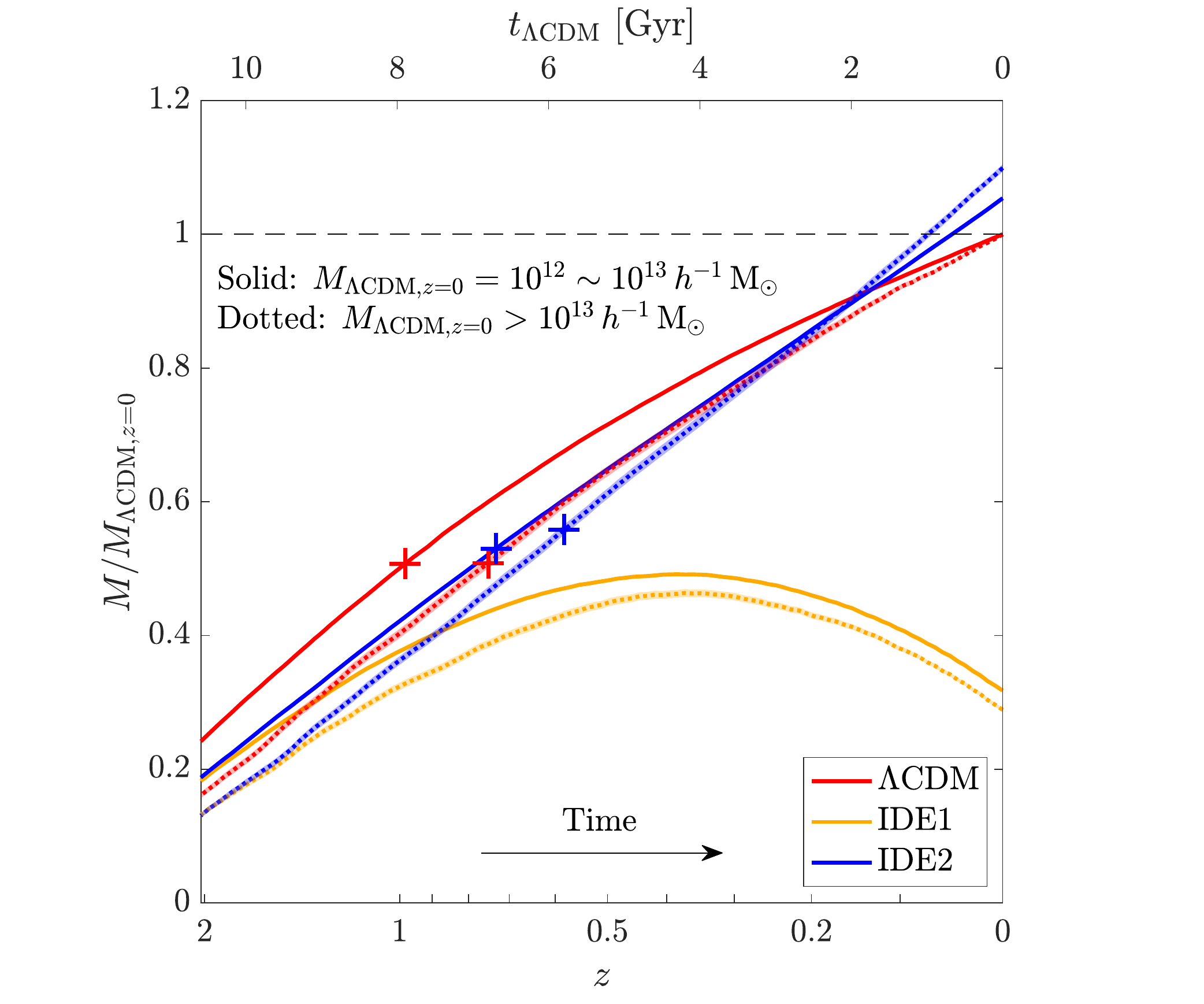}
    \includegraphics[width=\columnwidth]{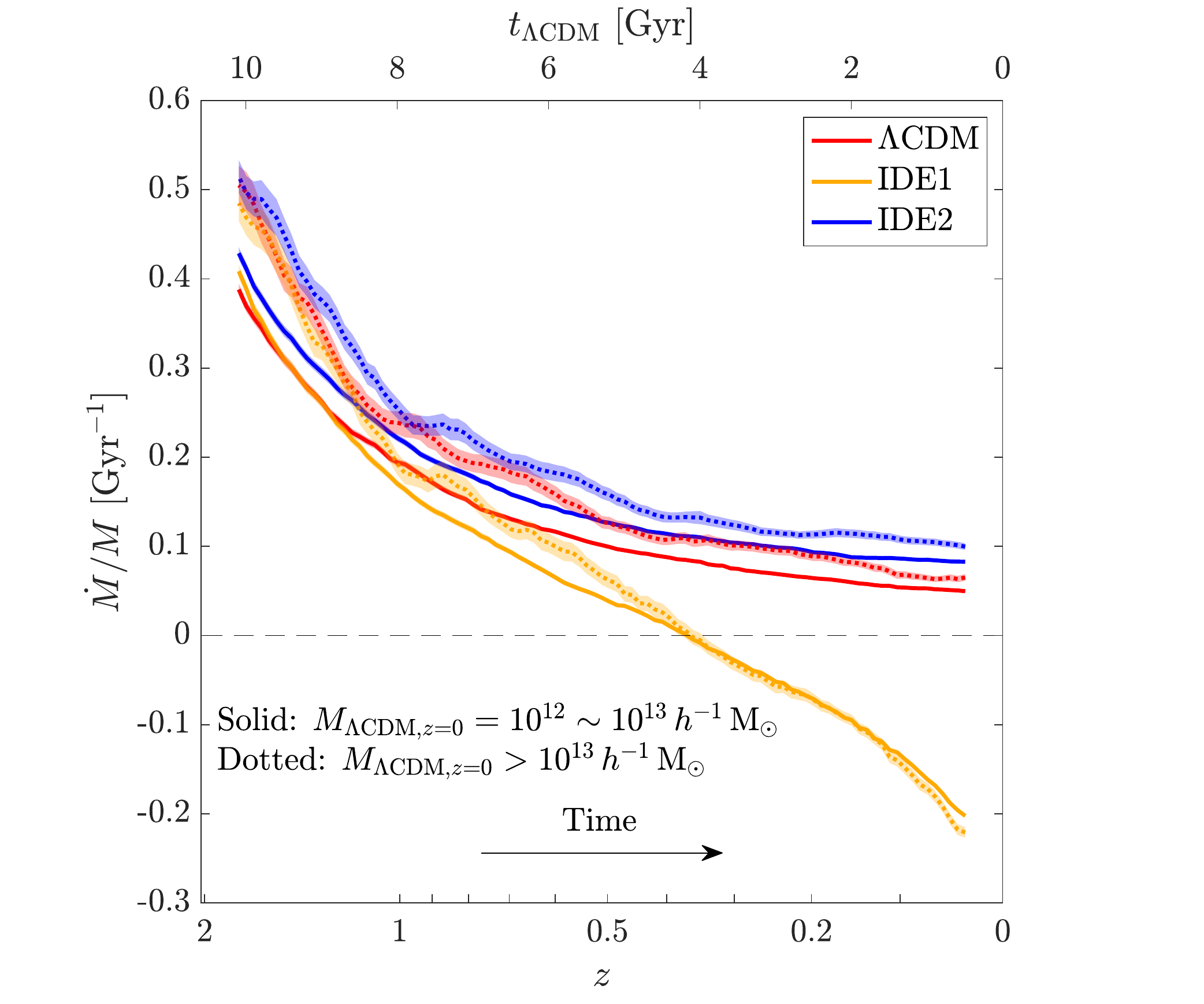}
    \caption{Top: average normalized mass growth history, $M(z)/M_{\Lambda\mathrm{CDM},z=0}$, of halos in $\Lambda$CDM (red), IDE1 (yellow) and IDE2 (blue) models. The solid and dotted lines show the low-mass and high-mass subsamples, respectively. The shaded region associated with each line represents the standard deviation of the means. The crosses mark the mean formation times. Bottom: similar to the top panel, but for the average relative mass growth rate, $\Dot{M}/M$.}
    \label{fig:History}
\end{figure}

From the halo mass functions shown above, we have learnt that the IDE models produce significantly different halo populations at low redshift comparing to the $\Lambda$CDM model. To offer a detailed picture of the growth of halos in the IDE models, in this subsection, we investigate the halo formation histories using the TRACE halo sample. We concentrate on the comparison of IDE1 and IDE2 with $\Lambda$CDM. The other two IDE models (i.e. IDE1$^\prime$ and IDE2$^\prime$) follow the similar trends but with milder differences to the $\Lambda$CDM model. In order to study the mass dependence, we divide the TRACE sample into two subsamples: the high-mass subsample of 346 halo triplets with $M_{\Lambda\mathrm{CDM},z=0}\geq10^{13}$\hMsun, and the low-mass subsample of 2664 halo triplets with $M_{\Lambda\mathrm{CDM},z=0}=10^{12}\sim10^{13}$\hMsun. Here $M_{\Lambda\mathrm{CDM},z=0}$ is the present-day halo mass in the $\Lambda$CDM model.

We normalize the mass of each halo, $M(z)$, by the corresponding $M_{\Lambda\mathrm{CDM},z=0}$. The average normalized mass growth vs. redshift for different models are shown in the top panel of Figure~\ref{fig:History}. With the halo growth history, we can compute the mean formation time of each subsample which is defined as the redshift when the main progenitor of a halo first reaches half of its halo mass at $z=0$. The mean formation redshifts for $\Lambda$CDM and IDE2 halos are marked with crosses in the top panel of Figure~\ref{fig:History}. For the $\Lambda$CDM model, more massive halos tend to form later, agreeing with many previous studies \citep[e.g.][]{Navarro1997ApJ...490..493N,Li2008MNRAS.389.1419L}. Similar trend can be found in the IDE2 model. Comparing to the $\Lambda$CDM counterparts, the masses of IDE2 halos are systematically lower at early epoch, but overtake at redshift $z\lesssim0.2$. This is consistent with the behavior of $\rho_{\rm m}(z)$ of IDE2, which is lower/higher than that of $\Lambda$CDM at high/low redshift.

In IDE1, the halo growth history is significantly different from that of the other two models. At $z\sim2$, being similar to the IDE2 halos, the IDE1 halos are less massive and grow slower comparing to the $\Lambda$CDM halos. At $z\sim1$, their growth becomes even slower and the halo masses are considerably lower than their counterparts of both $\Lambda$CDM and IDE2. After $z\sim 0.4$, the masses of IDE1 halos even turn around and start to decrease. Because of this unique evolution history, the aforementioned formation time is ill-defined in the IDE1 model, and we thus do not compute the mean formation time for this model.

At $z=0$, in comparison to the $\Lambda$CDM counterparts, IDE1 halos end with a mass ratio of $\sim0.3$ while IDE2 halos of $\sim1.1$. The mass difference between the $\Lambda$CDM model and the two IDE models is somewhat larger for the high-mass subsample than that of the low-mass subsample. This hints us that high-mass halos might help to put tighter constraints on IDE models.

We also analyse the mass growth rate for different models, which is defined as
\begin{equation}
    \Dot{M}(t)=\frac{M(t+0.5\Delta t)-M(t-0.5\Delta t)}{\Delta t},
	\label{eq:Mass growth rate}
\end{equation}
where we adopt $\Delta t=1\,\mathrm{Gyr}$. We test with $\Delta t=0.5\,\mathrm{Gyr}$ and $\Delta t=2\,\mathrm{Gyr}$, and the results are fairly similar. The calculated relative growth rates, $\Dot{M}/M$, are shown in the lower panel of Figure~\ref{fig:History}.

For $\Lambda$CDM, the $\Dot{M}/M$ decreases with cosmic time, reflecting that the growth of halos slows down. For the high-mass part, $\Dot{M}/M$ is systematically higher than that of the low-mass subsample. This is related to the fact that high-mass halos form later and their progenitors are typically in denser environments than the low-mass halos.

Similar $\Dot{M}/M$ behaviors are seen in IDE2 except that the $\Dot{M}/M$ is higher than that of $\Lambda$CDM. At $z\gtrsim1$, the $\Dot{M}$ are about the same for IDE2 and $\Lambda$CDM models as seen from the upper panel of Figure~\ref{fig:History}. However, the halo mass is systematically lower, and thus the $\Dot{M}/M$ is higher for IDE2. At the late epoch, because of the interaction between the two dark components that converts dark energy to dark matter, the halo growth is faster in IDE2, leading to a higher $\Dot{M}/M$ even the mass of IDE2 halos is larger at redshift $z\lesssim0.2$ than that of $\Lambda$CDM.

On the other hand, for IDE1, at $z\gtrsim1$, $\Dot{M}/M$ is about the same as that of $\Lambda$CDM for both high- and low-mass subsamples. This is because both $\Dot{M}$ and $M$ are lower in IDE1. At $z<1$, $\Dot{M}/M$ decreases much faster in IDE1 than the other two models, and becomes negative at $z\sim 0.4$. This is due to the interaction that dark matter decays into dark energy, which suppresses effectively the halo growth. At $z\lesssim0.4$, because of the decrease of the mass of dark matter particles, the gravitational potential of IDE1 halos gets shallower, and some of the halo member particles become unbound. Furthermore, they cannot accrete more particles. As a result, IDE1 halos are partially dissolved, and their mass decreases and thus the $\Dot{M}$ gets negative.

Our analyses here reveal clearly how the interaction between the two components affects the formation and evolution of dark matter halos. It is noted that the model parameters we adopt for IDE1, IDE2 and $\Lambda$CDM are the ones that can fit the observational data of both CMB and cosmic expansion history. However, their non-linear structure formation is significantly different, showing solidly the importance of including non-linear structure probes to tighten the constraints on different cosmological models.

\subsection{Halo density profile}
\label{sec:Density profile}

\begin{figure}
    \includegraphics[width=\columnwidth]{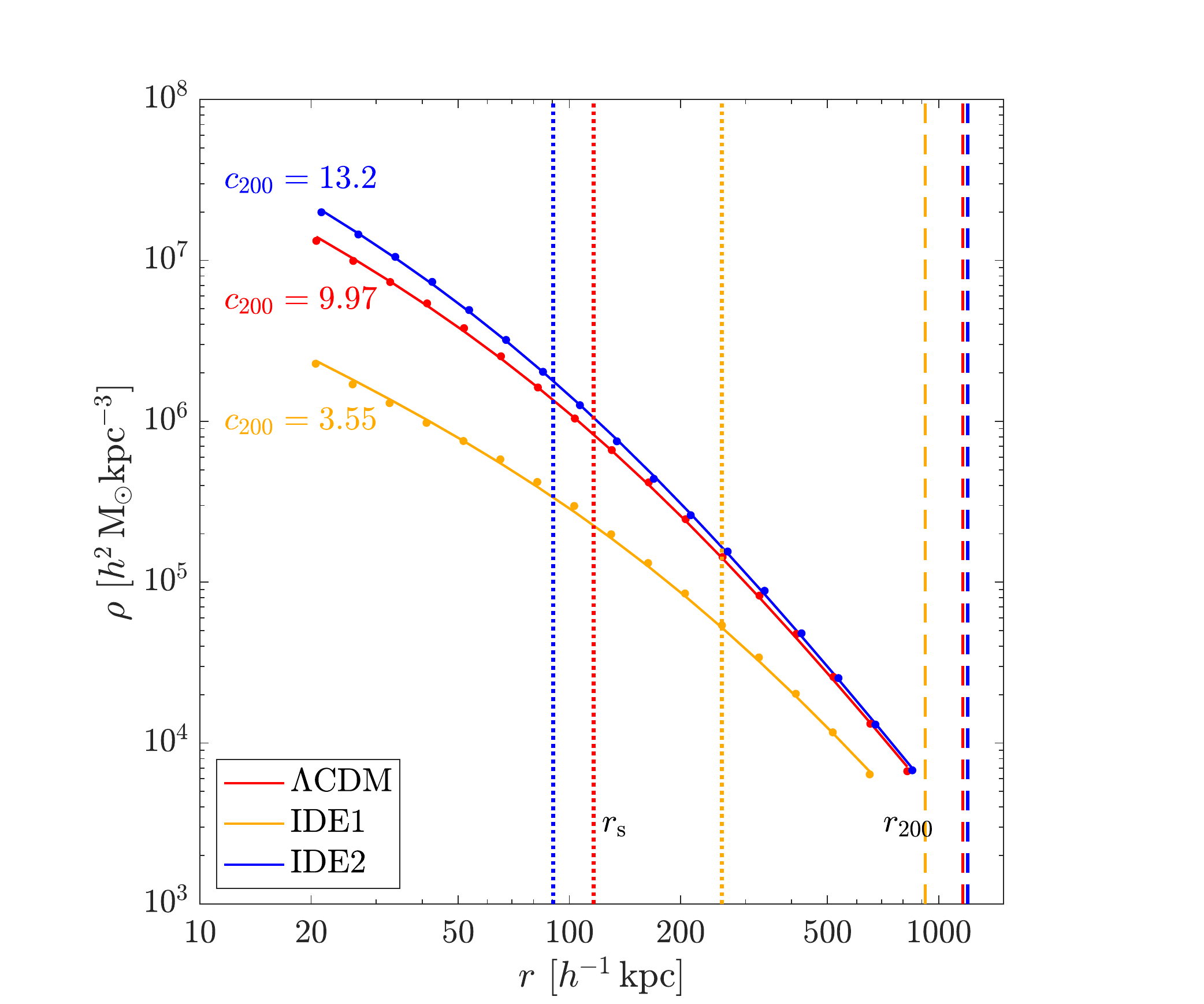}
    \caption{Density profiles of the $\Lambda$CDM halo (red) and its IDE1 (yellow) and IDE2 (blue) counterparts that shown in the bottom row of Figure~\ref{fig:LSS}. The data points are measured from simulations and the curve associated with each set of points is the corresponding NFW profile fitting. The best-fit concentrations are given explicitly. The dotted and dashed vertical lines mark the corresponding $r_{\mathrm{s}}$ and $r_{200}$, respectively.}
    \label{fig:Profile}
\end{figure}

\begin{figure}
    \includegraphics[width=\columnwidth]{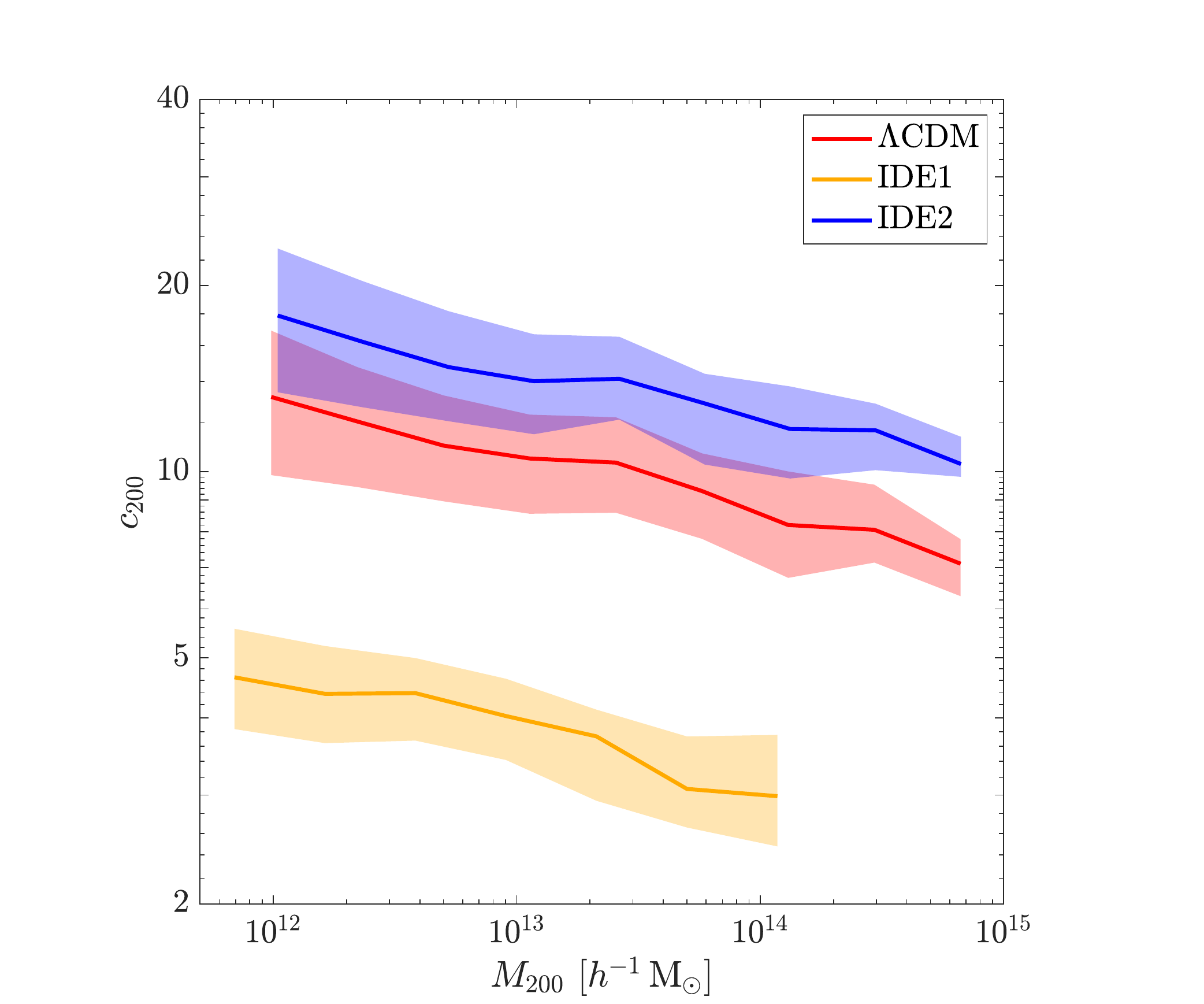}
    \caption{Concentration-mass relations of halos in $\Lambda$CDM (red), IDE1 (yellow) and IDE2 (blue) models. The lines show the medians and the shaded region associated with each line represents the range from 25 to 75 percentile.}
    \label{fig:cM}
\end{figure}

\begin{figure}
    \includegraphics[width=\columnwidth]{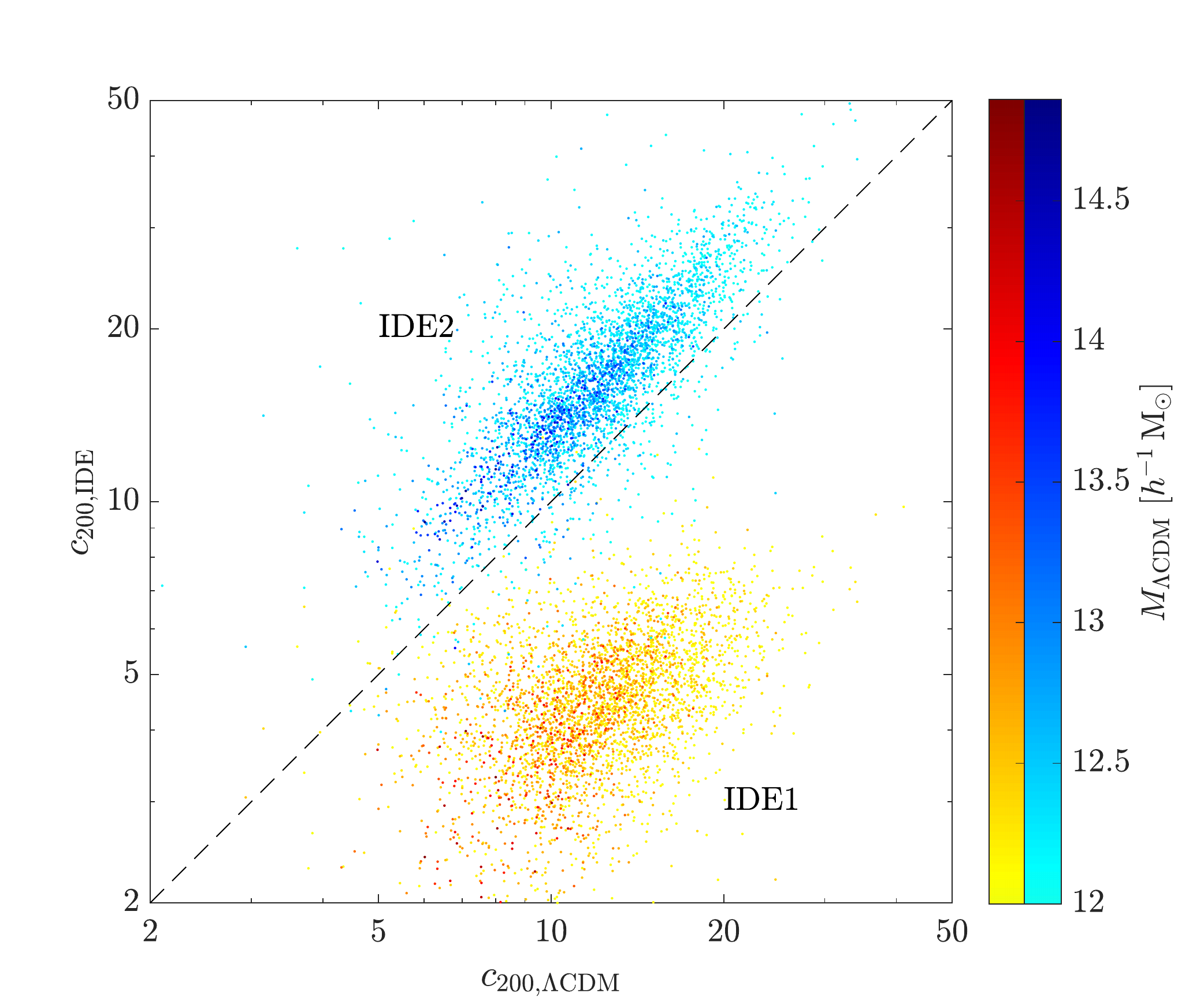}
    \caption{Contrast of halo concentrations between $\Lambda$CDM (horizontal axis) and IDE (vertical axis) counterparts. The yellow and blue groups show the results from IDE1 and IDE2, respectively. The color depth indicates the corresponding halo mass in $\Lambda$CDM.}
    \label{fig:cc}
\end{figure}

\begin{figure}
    \includegraphics[width=\columnwidth]{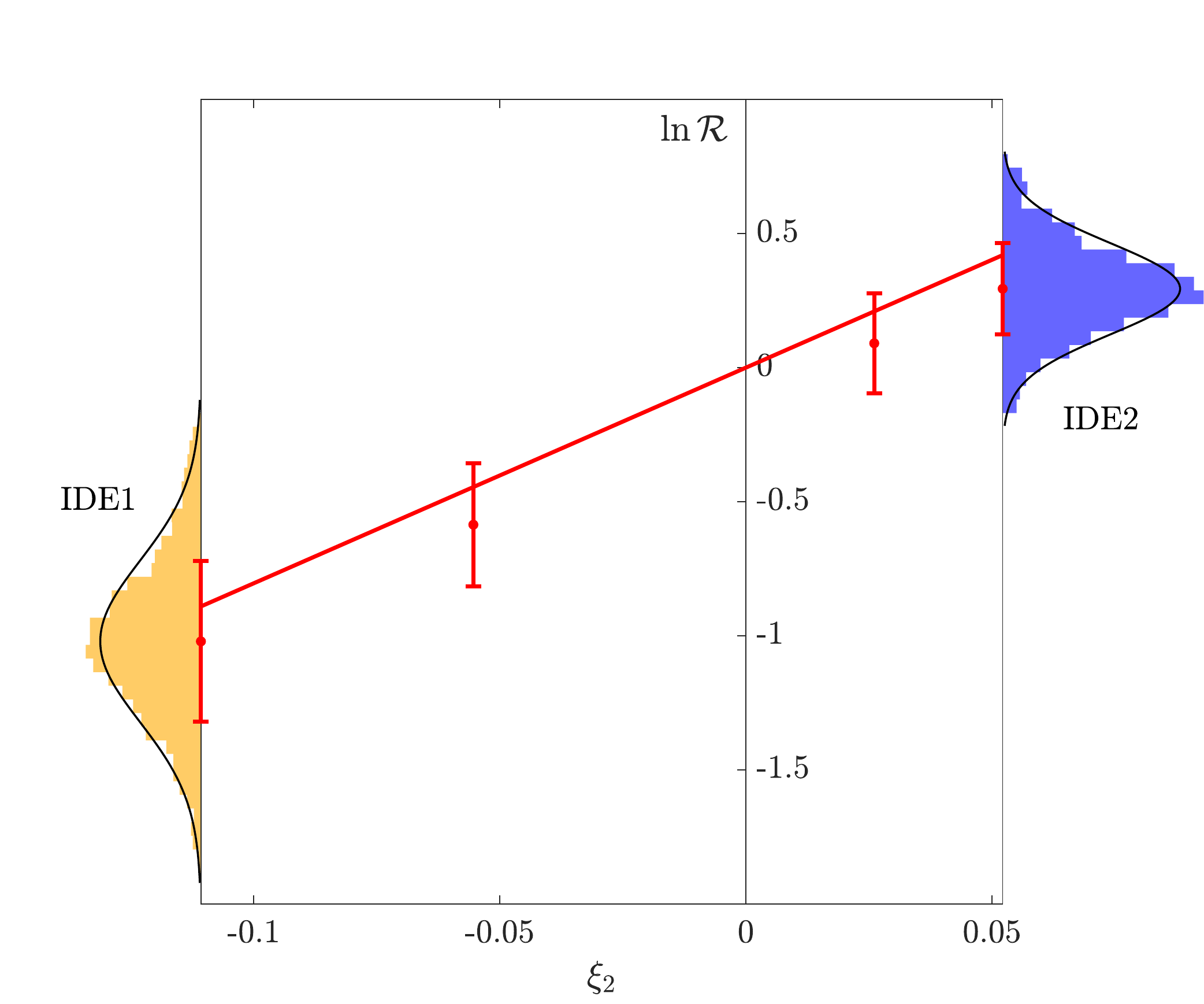}
    \caption{Correlation between the IDE-$\Lambda$CDM ratio of halo concentrations, $\mathcal{R}=c_{200,\mathrm{IDE}}/c_{200,\Lambda\mathrm{CDM}}$, and the interaction parameter, $\xi_{2}$. The red line shows the fitting model, $\ln \mathcal{R}=\alpha\xi_{2}$, where $\alpha=8.04\pm 0.03$. The four red data points with error bars show the means and $1\sigma$ errors of $\ln \mathcal{R}$ from IDE1, IDE1$^\prime$, IDE2$^\prime$ and IDE2, respectively. The distributions of $\ln \mathcal{R}$, and the corresponding normal fittings, from the best-fit IDE1 (yellow) and IDE2 (blue) are shown along the vertical axes.}
    \label{fig:ratio}
\end{figure}

In this subsection, we analyse the internal density profile of dark matter halos at $z=0$ for different models.

We measure the density profiles of the $z=0$ host halos with $\geq1000$ member particles in ALL sample, and all the halo triplets in MATCH sample. For each halo, we first divide the halo particles into 20 elementary bins logarithmically distributed in radial distance ranging from $0.01\,r_{200}$ to $r_{200}$. From these, we construct 18 overlapped bins each comprising 3 elementary bins to enrich the particle numbers. As in the Gadget code, the gravitational force only becomes exactly Newtonian beyond $2.8$ times the softening length (i.e. {$\sim 12$\hkpc} in our simulations), we drop those inner bins whose radii are smaller than 12 \hkpc.

We adopt the Navarro–Frenk–White (NFW) functional form \citep{Navarro1997ApJ...490..493N} in our analyses, which has been known to be an excellent description for the $\Lambda$CDM halo profile over a wide mass range. It is given by
\begin{equation}
    \rho(r)=\frac{\rho_{\mathrm{s}}}{\frac{r}{r_{\mathrm{s}}}\left(1+\frac{r}{r_{\mathrm{s}}}\right)^{2}},
	\label{eq:NFW}
\end{equation}
where $\rho(r)$ is the spherically averaged density at radius $r$. This is a two-parameter model: $r_{\mathrm{s}}$ is the scale radius and $\rho_{\mathrm{s}}$ is the characteristic density. The halo mass and radius are defined by
\begin{equation}
    M_{200}=200\rho_{\mathrm{m}}\cdot\frac{4}{3}\pi r_{200}^{3}=\int_{0}^{r_{200}}4\pi r^{2}\rho(r)\mathrm{d}r
	\label{eq:M200},
\end{equation}
and we further define the halo concentration as
\begin{equation}
    c_{200}=\frac{r_{200}}{r_{\mathrm{s}}}
	\label{eq:c},
\end{equation}
we then obtain
\begin{equation}
    \rho_{\mathrm{s}}=\rho_{\mathrm{m}}\cdot\delta=\rho_{\mathrm{m}}\cdot\frac{200}{3}\frac{c_{200}^{3}}{\ln{(1+c_{200})}-c_{200}/(1+c_{200})},
	\label{eq:rho_s}
\end{equation}
where $\delta$ is called the characteristic overdensity. Based on the equations above, we can calculate the average density, $\rho_{[R_{1},R_{2})}^{\mathrm{NFW}}$, within a radial bin $[R_{1},R_{2})$ where $R=r/r_{200}$. It can be written as
\begin{equation}
    \begin{split}
        \rho_{[R_{1},R_{2})}^{\mathrm{NFW}}=&\frac{200\rho_{\mathrm{m}}}{[\ln{(1+c_{200})}-c_{200}/(1+c_{200})]\cdot(R_{2}^{3}-R_{1}^{3})}\\
        &\cdot\left[\ln{\left(\frac{1+c_{200}R_{2}}{1+c_{200}R_{1}}\right)}+\frac{c_{200}(R_{1}-R_{2})}{(1+c_{200}R_{1})(1+c_{200}R_{2})}\right].
    \end{split}
	\label{eq:NFW_fitting}
\end{equation}
This is a one-parameter fitting formula with respect to $c_{200}$ for a halo profile with given $M_{200}$ (or $r_{200}$). We use the least-square method to do the fitting. Concretely, we define the following function to be minimized
\begin{equation}
    f(c)=\sum_{i=1}^{N_{\mathrm{bins}}}\left(\ln{\rho_{i}^{\mathrm{Simu.}}}-\ln{\rho_{i}^{\mathrm{NFW}}}\right)^{2},
	\label{eq:least square method}
\end{equation}
where $\rho_{i}^{\mathrm{Simu.}}$ is the density of the $i$th bin measured from a simulated halo. We do the NFW fitting for the halos from all five models. For the IDE models, the $\rho_{\mathrm{m}}$ is calculated by Equation~(\ref{eq:rho_m(a)}).

We find that the density profiles of the simulated dark matter halos in considered IDE models can be well fitted by the NFW profile for the mass range of our samples. Figure~\ref{fig:Profile} presents a typical example of the density profiles for a halo triplet in MATCH sample, where the three matched halos from IDE1, IDE2 and $\Lambda$CDM are the ones shown in the bottom row of Figure~\ref{fig:LSS}. The points are measured from simulated halos, and the lines are the fitted NFW profiles.

It is seen clearly that the IDE1 halo density profile markedly deviates from that of $\Lambda$CDM and IDE2 halos. It is overall lower because of the lowest mass of the IDE1 halo. Furthermore, it is significantly flatter than that of the other two models. This is closely related to the interaction that leads to dark matter decaying into dark energy. Thus the IDE1 halo cannot confine their member particles tightly. For IDE2, the profile is similar to that of $\Lambda$CDM, but somewhat steeper due to the increase of the dark matter content converted from dark energy.

In Figure~\ref{fig:Profile}, we also indicate the values of $r_{\mathrm{s}}$ and $r_{200}$ with vertical dotted and dashed lines, respectively, for the three halos. The IDE1 halo has the largest $r_{\mathrm{s}}$ and the smallest $r_{200}$, therefore its concentration is the smallest with $c_{200}=3.55$. For IDE2 halo, its profile is the steepest with the smallest $r_{\mathrm{s}}$, and the value of $r_{200}$ is about the same as that of the $\Lambda$CDM counterpart, resulting in the largest concentration of $c_{200}=13.2$. For the $\Lambda$CDM halo, $c_{200}=9.97$.

We emphasize that apart from the overall dimensional values of the halo density, its profile is independent of the dark matter particle mass in the halos. Thus by scaling the particle mass of IDE halos to that of the $\Lambda$CDM, we can only shift up (IDE1) and down (IDE2) the density distributions, but their profiles, e.g. the values of the concentration parameters, remain to be unchanged. Therefore the profile differences seen here indeed are the results of different halo formation histories among models.

In Figure~\ref{fig:cM}, we show the concentration-mass ($c_{200}$-$M_{200}$) relation at $z=0$ for the $\Lambda$CDM, IDE1 and IDE2 models, where the solid lines show the median values and the shaded regions are the range from 25 to 75 percentiles. It is seen that for all the three models, the concentration decreases with the halo mass. This is consistent with the cold dark matter scenario where low-mass halos form systematically earlier and thus more concentrate than the high-mass ones. We note that in \citet{Carlesi2014MNRAS.439.2943C}, they studied dark matter halo properties for coupled quintessence models in which dark matter converts to dark energy. They also observe lower halo concentrations than that of the $\Lambda$CDM. Our IDE1 results are qualitatively consistent with theirs although quantitatively different because of the different models.

What is significant from Figure~\ref{fig:cM} is that, the $c_{200}$-$M_{200}$ curves of the three models are nearly parallel in the log-log space, implying a nearly constant ratio in $c_{200}$ between IDE1 (IDE2) and $\Lambda$CDM independent of halo mass. We confirm this by analysing the MATCH sample. The results are shown in Figure~\ref{fig:cc}, where we contrast the concentration of each $\Lambda$CDM halo (horizontal axis) with that of its two IDE counterparts (vertical axis). Within each yellow (IDE1) or blue (IDE2) groups, the darker points are for more massive halos. Clearly, the ratio of the halo concentrations between IDE and $\Lambda$CDM counterparts, $\mathcal{R}=c_{200,\mathrm{IDE}}/c_{200,\Lambda\mathrm{CDM}}$, is indeed approximately independent of the halo mass.

For IDE1 and IDE2, the ratio $\mathcal{R}$ is very different. We expect it to depend on the interaction parameter, $\xi_{2}$. Considering a simple Taylor expansion in $\ln \mathcal{R}$ to the linear order, given $\ln \mathcal{R}=0$ at $\xi_{2}=0$, we have $\ln \mathcal{R}=\alpha\xi_{2}$. We fit this relation using the halo triplets in our MATCH sample. For enriching the data points, we further employ the IDE1$^\prime$ and IDE2$^\prime$ ones from the IDE1$^\prime$-$\Lambda$CDM-IDE2$^\prime$ triple-matched halo sample. The fitting is illustrated in Figure~\ref{fig:ratio}, where we plot along the left and right vertical axes the distributions of $\ln \mathcal{R}$ for IDE1 and IDE2 counterparts respectively, which can be well fitted by normal distributions. Similar distributions are seen for IDE1$^\prime$ and IDE2$^\prime$, which are not shown here. The red line is the fitting result with $\alpha=8.04\pm 0.03$, showing a positive correlation between $\ln \mathcal{R}$ and $\xi_{2}$. We see that $\ln \mathcal{R}$ is very sensitive to the interaction parameter $\xi_{2}$, thus it can act as a powerful probe to constrain IDE models by measuring dark matter halo profiles accurately. This again demonstrates the importance of non-linear structures in differentiating different cosmological models.

\subsection{Halo spin}
\label{sec:Spin}

\begin{figure*}
	\includegraphics[width=\columnwidth]{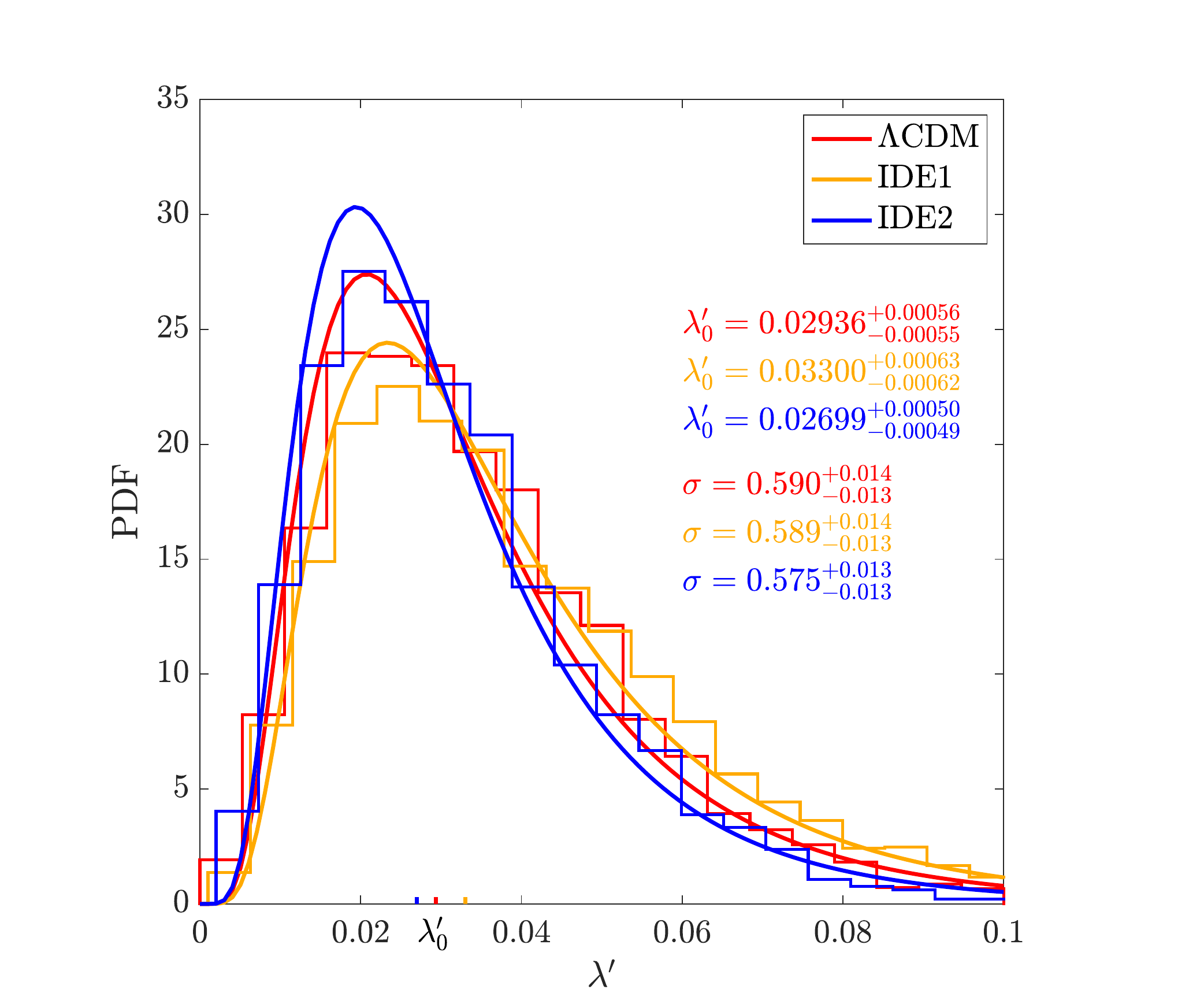}
	\includegraphics[width=\columnwidth]{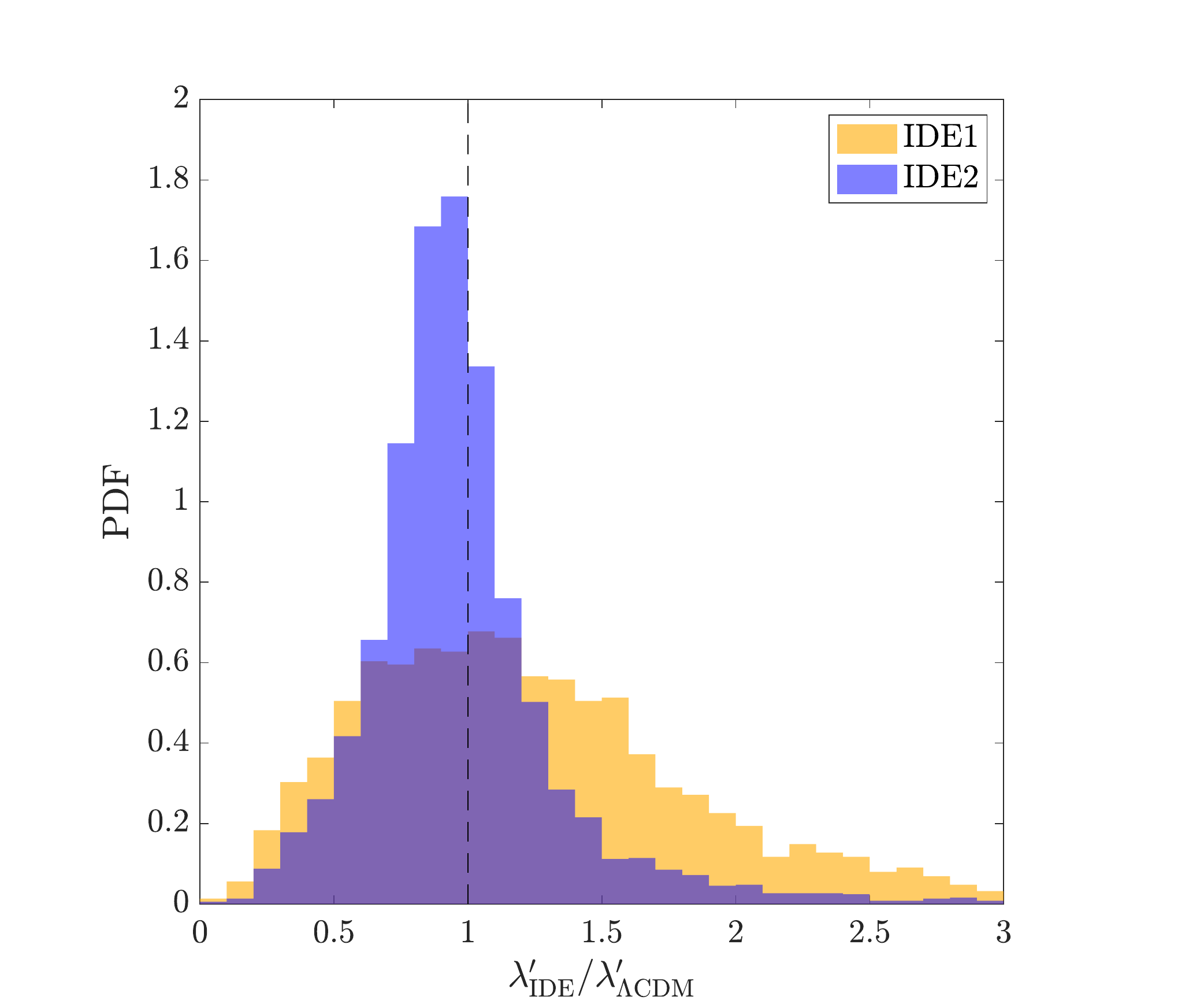}
    \caption{Left: distributions of halo spins in $\Lambda$CDM (red), IDE1 (yellow) and IDE2 (blue) models. The histograms are measured from simulations and the curve associated with each histogram is the corresponding log-normal fitting. The best-fit $\lambda_0^\prime$ and $\sigma$ are given explicitly. Right: distributions of the IDE-$\Lambda$CDM ratios of $\lambda^{\prime}$ from IDE1 (yellow) and IDE2 (blue). The vertical dashed line marks the ratio of $1$.}
    \label{fig:Spin}
\end{figure*}

The spin of dark matter halos is an important physical quantity that can affect directly the galaxy formation therein \citep*[e.g.][]{Mo1998MNRAS.295..319M}. Its origin is closely connected with the initial tidal torques \citep{Peebles1969ApJ...155..393P,Doroshkevich1970Afz.....6..581D,White1984ApJ...286...38W} and the halo formation history, particularly the merging history \citep{Maller2002MNRAS.329..423M,Vitvitska2002ApJ...581..799V,Peirani2004MNRAS.348..921P,DOnghia2007MNRAS.380L..58D}. Here we analyse halo spins for the IDE models. As before, we focus on the best-fit IDE1 and IDE2 models, and we note that the results from IDE1$^\prime$ and IDE2$^\prime$ are similar but with smaller differences from the $\Lambda$CDM model.

For each simulated halo, we measure its angular momentum using the following formula
\begin{equation}
    \boldsymbol{J}=m_{\mathrm{p}}\sum_{i=1}^{N_{\mathrm{p}}}(\boldsymbol{r}_{i}\times\boldsymbol{v}_{i}),
	\label{eq:Angular momentum}
\end{equation}
where $\boldsymbol{r}_{i}$ and $\boldsymbol{v}_{i}$ are the position and velocity of $i$th member particle in the centre-of-mass frame, respectively. The halo angular momentum is usually parameterized by the dimensionless spin parameter, $\lambda$, which was first proposed by \citet{Peebles1969ApJ...155..393P}, i.e.
\begin{equation}
    \lambda=\frac{J|E|^{1/2}}{GM^{5/2}},
	\label{eq:lambda}
\end{equation}
where $J$ is the magnitude of the total angular momentum, $E$ is the total energy and $M$ is the mass of a given halo. Dynamically, the spin parameter characterizes to what extent the structure of a halo is rotation-supported. The larger the $\lambda$ is, the more dominant the rotation component is; otherwise a halo with smaller $\lambda$ tends to be dominated more by the random motions. However, in practice, it is difficult to determine the total energy, $E$, of a realistic halo out of its complicated environment. Thus, effectively, \citet{Bullock2001ApJ...555..240B} have suggested an alternative spin parameter
\begin{equation}
    \lambda^{\prime}=\frac{J}{\sqrt{2}MVR}=\frac{j}{\sqrt{2GMR}},
	\label{eq:lambda'}
\end{equation}
where $j=J/M$ is the specific angular momentum and $V=\sqrt{GM/R}$ is the circular velocity at radius $R$. The parameter $\lambda^{\prime}$ is equivalent to $\lambda$ when we measure a fully relaxed halo within its virial radius. Here, we adopt the parameter $\lambda^{\prime}$ in our analyses.

Previous studies based on $\Lambda$CDM simulations show that $\lambda^{\prime}$ depends on halo mass only weakly \citep[see e.g.][]{Bett2007MNRAS.376..215B,Maccio2007MNRAS.378...55M,Knebe2008ApJ...678..621K}. We have investigated that in our IDE simulations and found similar conclusions. We therefore do not split the halos in a model into different mass bins here, but consider them as a whole. In $\Lambda$CDM, the distribution of $\lambda^{\prime}$ of a given halo population can be well fitted by a log-normal distribution
\begin{equation}
    P(\lambda^{\prime})=\frac{1}{\lambda^{\prime}\sqrt{2\pi}\sigma}\exp{\left(-\frac{\ln^{2}(\lambda^{\prime}/\lambda^{\prime}_{0})}{2\sigma^{2}}\right)},
	\label{eq:lognormal}
\end{equation}
where $\ln \lambda^{\prime}_{0}$ and $\sigma$ are, respectively, the mean and the standard deviation of $\ln\lambda^{\prime}$. The left panel of Figure~\ref{fig:Spin} presents the $\lambda^{\prime}$ distributions measured from our MATCH sample (histograms) and the corresponding log-normal fittings (solid lines). For $\Lambda$CDM halos, we have $\lambda^{\prime}_{0}=0.02936^{+0.00056}_{-0.00055}$, which is consistent with many previous studies \citep[e.g.][]{Bullock2001ApJ...555..240B,Trowland2013ApJ...762...72T}. The IDE halo spins in our samples can also be well fitted by Equation~(\ref{eq:lognormal}). In comparison to $\Lambda$CDM, we find a model-dependent shift of $\lambda^{\prime}$ for IDE models with IDE1 having a larger $\lambda^{\prime}_{0}$, and IDE2 a smaller one. Meanwhile, the $\sigma$ for three models are very close considering the errors. As a detailed comparison, in the right panel of Figure~\ref{fig:Spin}, we show the distributions of the IDE-$\Lambda$CDM ratios of $\lambda^{\prime}$ from MATCH sample. The median of $\lambda^{\prime}_{\mathrm{IDE1}}/\lambda^{\prime}_{\Lambda\mathrm{CDM}}$ is $\sim 1.17$, and that of $\lambda^{\prime}_{\mathrm{IDE2}}/\lambda^{\prime}_{\Lambda\mathrm{CDM}}$ is $\sim 0.93$.

In comparison to the $\Lambda$CDM halos, the larger (smaller) $\lambda^{\prime}$ of IDE1 (IDE2) counterparts indicates that the rotation component dominates more (less) for the total motions of member particles. This can be understood from two aspects. On the one hand, IDE1 (IDE2) halos have smaller (larger) velocity dispersions than that of the $\Lambda$CDM counterparts given their shallower (deeper) potential wells as discussed in section~\ref{sec:Density profile}. On the other hand, as described in section~\ref{sec:Simulations}, there is an additional acceleration for the simulated particles in IDE models arising from the dark energy-dark matter interaction. It acts as accelerating (decelerating) the particles in IDE1 (IDE2) model, leading to an increase (decrease) of the specific angular momentum $j$. The combination of the two effects leads to the larger (smaller) values of $\lambda^{\prime}$ for IDE1 (IDE2) halos in comparison with that of $\Lambda$CDM.

\subsection{Halo shape}
\label{sec:Shape}

\begin{figure}
    \includegraphics[width=\columnwidth]{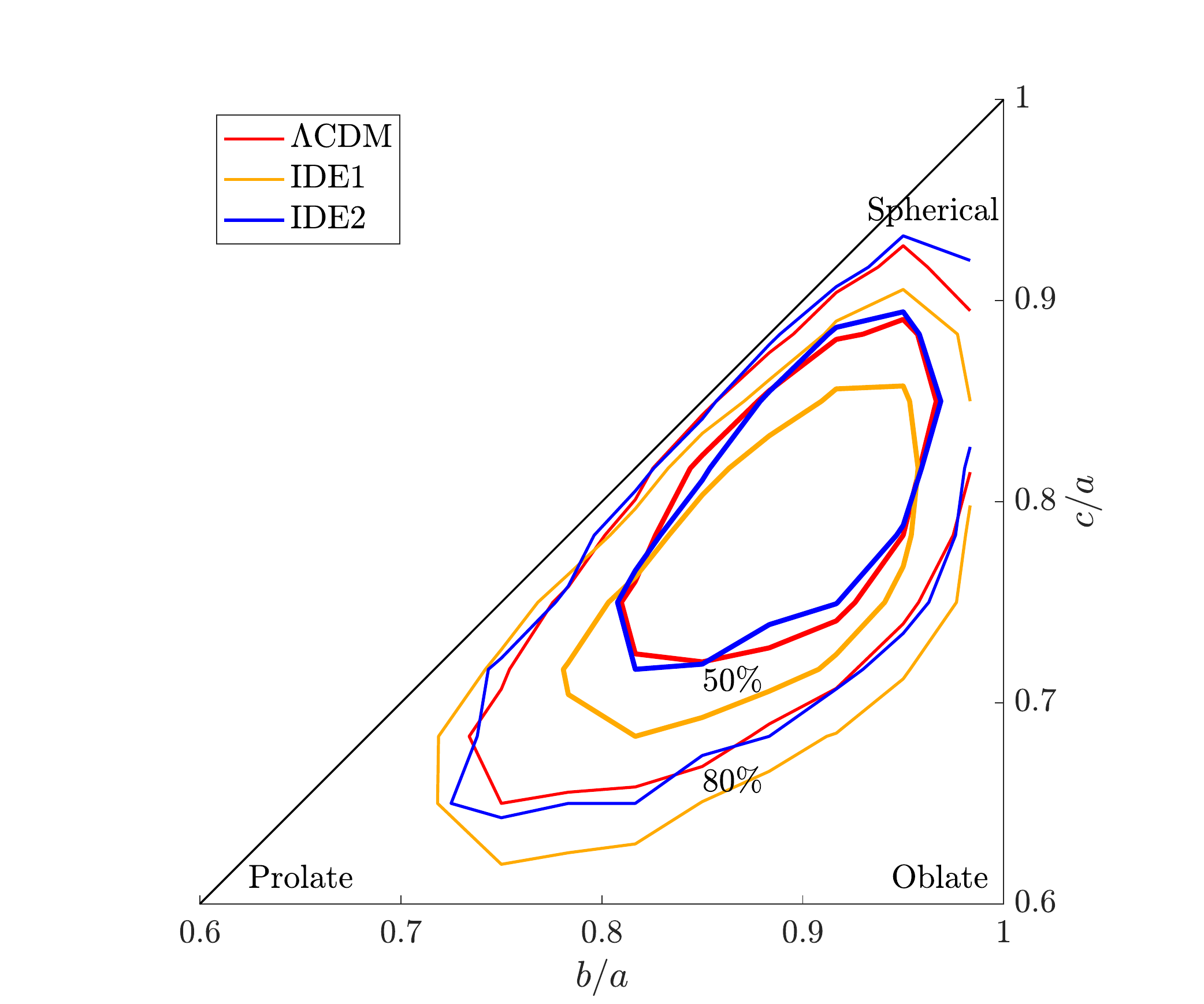}
    \caption{Distributions of halo shapes in $\Lambda$CDM (red), IDE1 (yellow) and IDE2 (blue) models in a $b/a\ \times\ c/a$ parameter space. The inner (thick) and outer (thin) lines show the $50\%$ and $80\%$ probability contours, respectively.}
    \label{fig:Shape_bc}
\end{figure}

\begin{figure}
	\includegraphics[width=\columnwidth]{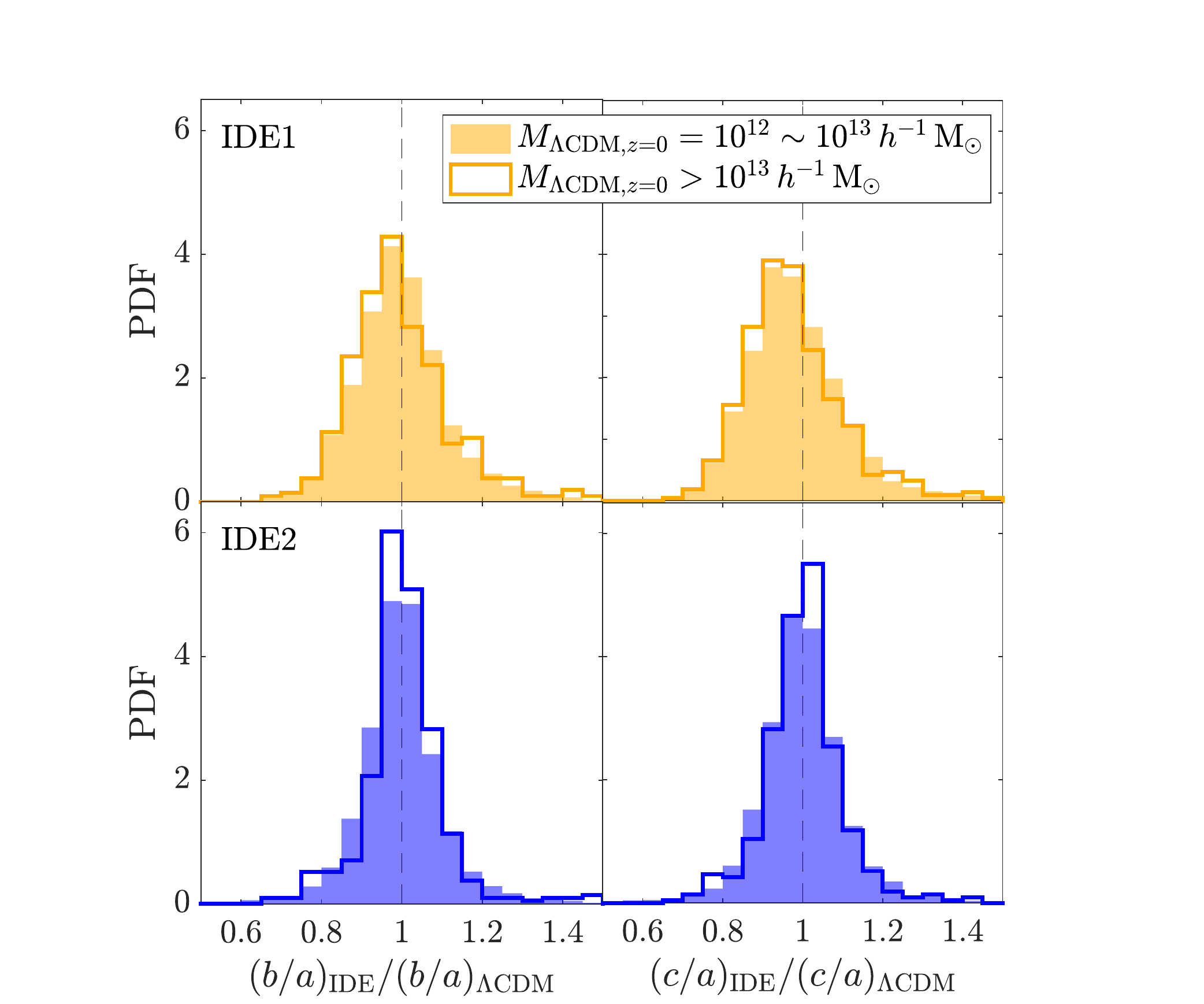}
    \caption{Distributions of the IDE-$\Lambda$CDM ratios of $b/a$ (left column) and $c/a$ (right column) from IDE1 (top row) and IDE2 (bottom row), respectively. The filled and open histograms show the low-mass and high-mass subsamples, respectively. The vertical dashed lines mark the ratio of $1$.}
    \label{fig:Shape_PDF}
\end{figure}

Here we further analyse the shape of dark matter halos in different models. It is a non-trivial task to model the detailed shape of a realistic halo \citep[see e.g.][]{Jing2002ApJ...574..538J}. In this study, following many literature on halo shapes, we model the dark matter halo as a 3-D ellipsoid, which is characterized by its three orthogonal principal axes: $a$ (major axis), $b$ (intermediate axis) and $c$ (minor axis) satisfying $a\geq b\geq c$.

The three principle axes of a simulated halo can be computed from the inertia tensor, $\mathbf{I}$, which relates the angular momentum $\boldsymbol{J}$ and angular velocity $\boldsymbol{\omega}$ as $\boldsymbol{J}=\mathbf{I}\boldsymbol{\omega}$. The component of $\mathbf{I}$ can be written as
\begin{equation}
    I_{\alpha\beta}=m_{\mathrm{p}}\sum_{i=1}^{N_{\mathrm{p}}}(|\boldsymbol{r}_{i}|^{2}\delta_{\alpha\beta}-r_{i,\alpha}r_{i,\beta}),
	\label{eq:Inertia tensor}
\end{equation}
where $\boldsymbol{r}_{i}$ is the position of $i$th member particle with respect to the halo centre, $\alpha$ and $\beta$ are the spatial tensor indices with value of 1, 2 or 3, and $\delta_{\alpha\beta}$ is the Kronecker Delta function with $\delta_{\alpha\beta}=1$ for $\alpha=\beta$ and 0 otherwise. We diagonalize the tensor to obtain the lengths of the three principal axes, which are given by
\begin{equation}
    \mathbf{I}=\frac{1}{5}M\begin{pmatrix}
    b^{2}+c^{2} & 0 & 0 \\
    0 & a^{2}+c^{2} & 0 \\
    0 & 0 & a^{2}+b^{2} 
    \end{pmatrix},
	\label{eq:Eigenvalues}
\end{equation}
where $M$ is the mass of considered halo.

We then calculate the axial ratios, $b/a$ and $c/a$, for each halo in the MATCH sample. The probability distribution contours in $b/a\ \times\ c/a$ parameter space for the $\Lambda$CDM, IDE1 and IDE2 counterparts are shown in Figure~\ref{fig:Shape_bc}, where the top-right corner indicates spherical-like structures, and the bottom-left and -right are for prolate and oblate ones, respectively. We can see that for IDE1, the contours shift systematically toward the bottom and to the left, showing that IDE1 halos tend to be more aspherical systematically than the ones in the other two models. For IDE2 halos, their axial ratio distributions are about the same as that of $\Lambda$CDM.

We also quantitatively confirm the above results by computing the IDE-$\Lambda$CDM ratios for $b/a$ and $c/a$, as shown in Figure~\ref{fig:Shape_PDF}. For IDE1, the medians are $\sim 0.98$ and $\sim 0.96$ for $b/a$ and $c/a$ ratios, respectively. For IDE2, both ratios are close to $\sim1$ based on our sample. For IDE1$^\prime$ and IDE2$^\prime$, the corresponding ratios are closer to 1. To be consistent with the results of halo spin, for IDE1 halos, the additional acceleration of member particles arising from the dark energy-dark matter interaction prevents the halo relaxation and increases the halo anisotropy. Again, we observe a weak mass dependence from the filled and open histograms for low- and high-mass subsamples respectively in Figure~\ref{fig:Shape_PDF}. The high-mass subsample of IDE1 shows a slightly larger deviation from that of the $\Lambda$CDM than the low-mass subsample.

We further calculate the triaxiality parameter introduced by \citet{Franx1991ApJ...383..112F} for the halos in MATCH sample, defined as
\begin{equation}
    T=\frac{a^{2}-b^{2}}{a^{2}-c^{2}},
	\label{eq:Triaxiality}
\end{equation}
which reveals more clearly how prolate ($T\sim1$) or oblate ($T\sim0$) a halo is than the axial ratios. The medians and 25 to 75 percentiles for different models are: $T_{\Lambda\mathrm{CDM}}=0.597^{+0.147}_{-0.168}$, $T_{\mathrm{IDE1}}=0.581^{+0.155}_{-0.177}$ and $T_{\mathrm{IDE2}}=0.602^{+0.155}_{-0.182}$. The results show that in all the models, the dark matter halos are dominantly triaxial-like, and there is little difference for the triaxiality parameter between $\Lambda$CDM and IDE models considering the errors.

\section{Conclusions}
\label{sec:Conclusions}

Based on a set of self-consistent IDE and $\Lambda$CDM $N$-body simulations, we systematically analyse the formation history of dark matter halos and their properties in IDE models and compare them with the $\Lambda$CDM counterparts. We aim to understand how the interaction between the two dark components in the Universe affects the non-linear cosmic structure formation, and thus to find sensitive probes in constraining different cosmological models.

We investigate four IDE models, IDE1, IDE1$^\prime$, IDE2 and IDE2$^\prime$. All models have the dark energy-dark matter interactions proportional to the density of dark energy, but the signs of the interaction parameter $\xi_2$ are different. For IDE1 and IDE1$^\prime$, dark matter decays into dark energy while the opposite for IDE2 and IDE2$^\prime$. The IDE1 and IDE2 employ the best-fit parameters from previous constraints, and the ones marked with prime symbol employ the parameters of midpoint values between the $\Lambda$CDM and corresponding best-fit models.

We summarize our main results from IDE1, IDE2 and their comparisons with $\Lambda$CDM as follows. The IDE1$^\prime$ and IDE2$^\prime$ behave similarly to the IDE1 and IDE2, respectively, but with smaller differences from $\Lambda$CDM.

\begin{enumerate}
 \item The interaction of dark energy and dark matter affects the halo mass function significantly. In particular, for IDE1 in which dark matter decays into dark energy, the structure formation is slowed down, and the halo mass function is markedly lower than that of $\Lambda$CDM at low redshift. On the contrary, we observe a faster increase of the halo mass function from high to low redshift in IDE2 model where dark energy transfers to dark matter. The sensitive dependence of the halo mass function on the dark energy-dark matter interaction provides a powerful non-linear probe in constraining IDE models.
 
 \item The halo formation history shows distinctly different behaviors in IDE models in comparison with $\Lambda$CDM model. The IDE2 halos have larger mass growth rates $\Dot{M}$ and also the relative mass growth rates $\Dot{M}/M$ at low redshift. At $z=0$, the mass ratio of halos in IDE2 to their $\Lambda$CDM counterparts is $\sim1.1$. For IDE1 halos, their growth is greatly suppressed at $z\lesssim 1$, and even turns to be negative at $z\lesssim 0.4$. As a result, at $z=0$, the mass of IDE1 halos is only about 0.3 times that of the $\Lambda$CDM counterparts. The mass contrast between IDE models and $\Lambda$CDM model shows a weak mass dependence with the high-mass subsample having a larger mass difference than the low-mass ones.

 \item For the internal density profile of dark matter halos, they all can be well fitted by the NFW functional form for both IDE and $\Lambda$CDM models. However, the characteristic halo parameters are very different for different models. Comparing to the $\Lambda$CDM counterparts, the IDE1 halos have lower concentrations and internal potentials, and the IDE2 halos show the opposite. Notably, we find that the IDE-$\Lambda$CDM ratio of halo concentrations, $\mathcal{R}=c_{200,\mathrm{IDE}}/c_{200,\Lambda\mathrm{CDM}}$, is nearly independent of halo mass. By fitting a simple linear model in $\ln \mathcal{R}$, i.e. $\ln \mathcal{R}=\alpha\xi_2$, we obtain $\alpha=8.04\pm 0.03$. This high value of $\alpha$ reflects a sensitive dependence of the halo internal density profile on the interaction parameter, offering an important means to constrain IDE models.
 
 \item We also analyse the halo spin and shape for different models by measuring the specific spin parameter $\lambda^{\prime}$, the axial ratios $b/a$ and $c/a$, and the triaxiality parameter $T$. To compare with the $\Lambda$CDM counterparts, we find that IDE1 halos have systematically larger $\lambda^{\prime}$, smaller $b/a$ and $c/a$, which indicate the structures of IDE1 halos are more anisotropic, and dominated more by the overall rotations. IDE2 halos have smaller $\lambda^{\prime}$ than the $\Lambda$CDM counterparts, but the shape parameters are about the same. For all models, their halos are primarily triaxial, and the $T$ parameters for the halos in $\Lambda$CDM and IDE models are about the same.

\end{enumerate}

Our results quantitatively show the impacts of the interaction between dark energy and dark matter on non-linear structure formation. As the observed halo mass function and $c_{200}$-$M_{200}$ relation are largely consistent with the predictions from $\Lambda$CDM simulations \citep{Tinker2008ApJ...688..709T,Du2015ApJ...814..120D,Xu2021ApJ...922..162X}, we can safely rule out the considered best-fit parameter set of IDE1 although which can fit CMB and cosmic expansion history data well. This clearly shows the indispensable role of non-linear probes in cosmological studies. The approximately constant ratio of the $\mathcal{R}=c_{200,\mathrm{IDE}}/c_{200,\Lambda\mathrm{CDM}}$ and its dependence on the interaction parameter promise a sensitive means to constrain IDE model parameters by measuring dark matter halo density profiles. The quantitative constraints based on our results and together with the observational data will be carried out in our future work.

In addition to the simulations of the five models presented in previous sections, we also run and analyse two toy IDE models with their cosmological parameters being the same as that of the $\Lambda$CDM model except $\xi_2=-0.05$ and $w_\mathrm{d}=-0.999$ for the IDE1-toy model and $\xi_2=0.05$ and $w_\mathrm{d}=-1.001$ for the IDE2-toy model. Because of the matter density being normalized to the present observed value, the IDE1-toy model/IDE2-toy model has higher/lower $\Omega_{\rm m} (z)$ than that of $\Lambda$CDM, in contrast to the IDE models considered in the main text. By analysing the toy models, we find that while the quantitative results do depend on the cosmological parameter settings, qualitatively, all the IDE1-like models with dark matter transferring to dark energy show similar behaviors in slowing down the structure formation. On the other hand, for the IDE2-like models with dark energy converting to dark matter, the structure formation is getting strengthened toward the present time.

Limited by simulation resolutions, our investigations here focus mainly on host halos with mass in $\Lambda$CDM model above $10^{12}$\hMsun. To further extend to smaller halos, simulations of higher resolutions are needed. We also note that our simulations are dark-matter-only without including baryons. Although we do not expect qualitative changes about our conclusions regarding IDE1 and IDE2 in comparison with that of $\Lambda$CDM, detailed studies involving baryon physics are desirable in order to better confront with observations above and below galactic scales.

\section*{Acknowledgements}

This work was inspired by the discussions during the 2nd HOUYI Workshop for Non-standard Cosmological Models in Kunming, China, 2019. The calculations of this study were partly done on the Yunnan University Astronomy Supercomputer. YL acknowledges the supports from the Research $\&$ Innovation Project for Postgraduates of Yunnan University (No. 2019z060) and from NSFC of China (No. 11973036). SHL acknowledges the support by the European Research Council via ERC Consolidator Grant KETJU (No. 818930). ZHF and XKL are supported by NSFC of China under Grant No. 11933002 and No. U1931210, and a grant from the CAS Interdisciplinary Innovation Team. XKL also acknowledges the supports from NSFC of China under Grant No. 11803028 and No. 12173033, YNU Grant No. C176220100008, and the research grants from the China Manned Space Project with No. CMS-CSST-2021-B01. ZHF also acknowledges the supports from NSFC of China under No. 11653001, and the research grants from the China Manned Space Project with No. CMS-CSST-2021-A01. JJZ was supported by IBS under the project code, IBS-R018-D1, and the research grants from the China Manned Space Project with No. CMS-CSST-2021-A03. RA acknowledges the support from National Science Foundation under Grant No. PHY-2013951 at USC.

\section*{Data Availability}

The simulation data used in this article will be shared on reasonable request to the authors.



\bibliographystyle{mnras}
\bibliography{liuyun} 








\bsp	
\label{lastpage}
\end{document}